\documentclass[traditabstract]{aa}

\usepackage{graphicx}
\usepackage{natbib}
\bibpunct{(}{)}{;}{a}{}{,} % to follow the A&A style

\usepackage{amsmath,amssymb}
\usepackage{url}
\usepackage{fixltx2e}

\usepackage{txfonts}

\usepackage{color}
\def\changed{}

\newcommand{\msun}{\ensuremath{\mathrm{M}_{\odot}}}

\newcommand{\lsun}{\ensuremath{\mathrm{L}_{\odot}}}

\newcommand{\msunperyr}{\ensuremath{\msun\,\mathrm{yr}^{-1}}}

\renewcommand{\deg}{\ensuremath{^\circ}}

\begin{document}

\title{Wind bubbles within H\,\textsc{ii} regions around slowly moving stars}
\author{Jonathan~Mackey\inst{1} \and
        Vasilii~V.~Gvaramadze\inst{2,3,4} \and
        Shazrene~Mohamed\inst{5} \and
        Norbert~Langer\inst{1}}
\offprints{JMackey@astro.uni-bonn.de}
\institute{
  Argelander-Institut f\"ur Astronomie, Auf dem H\"ugel 71, 53121 Bonn, Germany
  \and
  Sternberg Astronomical Institute, Lomonosov Moscow State University, Universitetskij Pr.~13, Moscow 119992, Russia
  \and
  Isaac Newton Institute of Chile, Moscow Branch, Universitetskij Pr.\ 13, Moscow 119992, Russia
  \and
  Space Research Institute, Russian Academy of Sciences, Profsoyuznaya 84/32, 117997 Moscow, Russia
  \and
  South African Astronomical Observatory, P.O.\ box 9, 7935 Observatory, South Africa
}

\date{Received DD Month 2014 / Accepted DD Month 2014}

\abstract{
  Interstellar bubbles around O stars are driven by a combination of the star's wind and ionizing radiation output.
  The wind contribution is uncertain because the boundary between the wind and interstellar medium is difficult to observe.
  Mid-infrared observations (e.g., of the H\,\textsc{ii} region RCW\,120) show arcs of dust emission around O stars, contained well within the H\,\textsc{ii} region bubble.
  These arcs could indicate the edge of an asymmetric stellar wind bubble, distorted by density gradients and/or stellar motion.
  We present two-dimensional, radiation-hydrodynamics simulations investigating the evolution of wind bubbles and H\,\textsc{ii} regions around massive stars moving through a dense ($n_\mathrm{H}=3000\,\mathrm{cm}^{-3}$), uniform medium with velocities ranging from $4$ to $16\,\mathrm{km}\,\mathrm{s}^{-1}$.
  The H\,\textsc{ii} region morphology is strongly affected by stellar motion, as expected, but the wind bubble is also very aspherical from birth, even for the lowest space velocity considered.
  Wind bubbles do not fill their H\,\textsc{ii} regions (we find filling factors of 10-20 per cent), at least for a main sequence star with mass $M_\star\sim30\,\msun$.
  Furthermore, even for supersonic velocities the wind bow shock does not significantly trap the ionization front.
  X-ray emission from the wind bubble is soft, faint, and comes mainly from the turbulent mixing layer between the wind bubble and the H\,\textsc{ii} region.
  The wind bubble radiates $<1$ per cent of its energy in X-rays; it loses most of its energy by turbulent mixing with cooler photoionized gas.
  Comparison of the simulations with the H\,\textsc{ii} region RCW\,120 shows that its dynamical age is $\lesssim0.4$ Myr and that stellar motion $\lesssim4\,\mathrm{km}\,\mathrm{s}^{-1}$ is allowed, implying that the ionizing source is unlikely to be a runaway star but more likely formed in situ.
  The region's youth, and apparent isolation from other O or B stars, makes it very interesting for studies of massive star formation and of initial mass functions.
}

%\authorrunning{}
%\titlerunning{}
\maketitle

%%% ------------------------------------------------------------
%%% ------------------------------------------------------------
\section{Introduction}
\label{sec:intro}
%%% ------------------------------------------------------------
%%% ------------------------------------------------------------

Observations of interstellar bubbles along the Galactic plane in the mid-infrared (mid-IR) \citep{DehSchAndEA10, KenSimBreEA12, SimPovKenEA12} show that the interiors of H\,\textsc{ii} regions contain dust, and that many of them have arcs of 24 $\mu$m dust emission near the central ionizing star (or stars), well within the H\,\textsc{ii} region border.
\citet{OchVerCoxEA14} interpret these arcs (based on the original idea by \citealt{VanMcC88}) as emission from dust grains that have decoupled from the gas and are deflected away from the ionizing stars by their radiation pressure.
An alternative interpretation -- that the arcs delineate the edge of a stellar wind bubble within a larger H\,\textsc{ii} region -- has not so far been explored with multidimensional simulations.
One-dimensional calculations \citep{PavKirWie13} show that the interpretation of the arcs is complicated by dust processing within H\,\textsc{ii} regions, but that it is feasible that they represent the boundary between stellar wind and the interstellar medium (ISM).

\begin{figure*}
\centering
\resizebox{0.8\hsize}{!}{\includegraphics{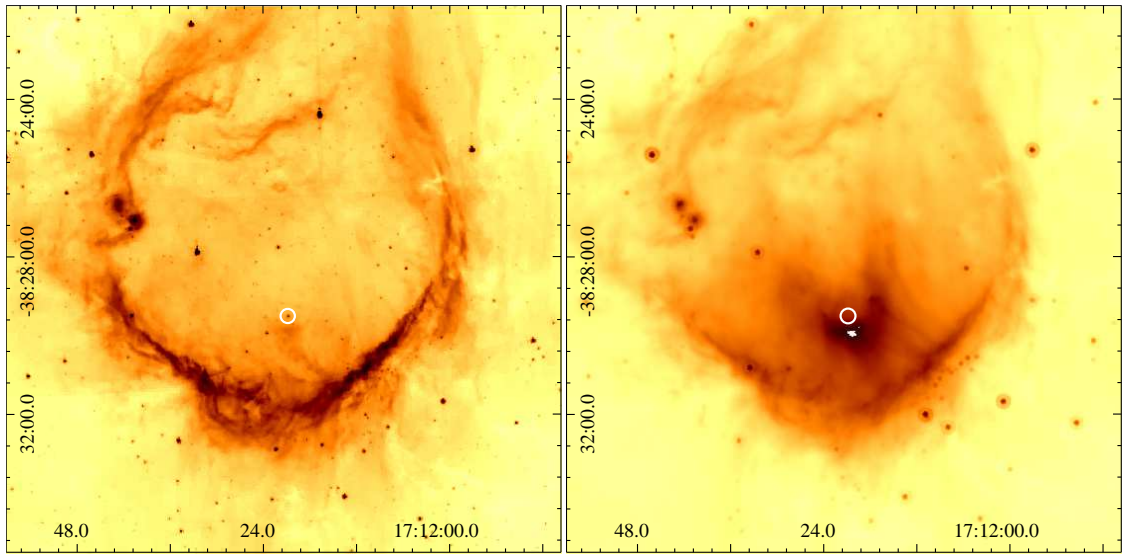}}
\caption{
  The Galactic H\,\textsc{ii} region RCW\,120 in the mid-IR from \emph{Spitzer}  8\,$\mu$m  (left), and  24\,$\mu$m (right).
  The white circle shows the location of the ionizing star CD\,$-$38$\deg$11636 that drives the nebula's expansion.
  The coordinates are in units of RA(J2000) and Dec.(J2000) on the horizontal and
vertical scales, respectively.
  The outer ring, clearest at 8\,$\mu$m, is the H\,\textsc{ii} region boundary and the inner arc seen at 24\,$\mu$m may show the edge of the stellar wind bubble.
  }
\label{fig:rcw120}
\end{figure*}

One of the best examples of an interstellar bubble with a 24 $\mu$m arc is \object{RCW 120}, shown in Fig.~\ref{fig:rcw120}.
It is a Galactic H\,\textsc{ii} region bounded by a massive, dense shell with mass, $M_\mathrm{sh}\approx1200-2100\,\msun$ \citep{DehZavSchEA09}, embedded in a molecular cloud.
\citet{ZavPomDehEA07} estimate the number density of atoms ($n_0$) in the surrounding ISM to be $n_0\approx1400-3000\,\mathrm{cm}^{-3}$.
The nebula is ionized by the star CD\,$-$38$\deg$11636, an O6-8\,V/III star with $M_\star\approx30\,\msun$, which could be a double star \citep{MarPomDehEA10}.
It is almost certainly a main sequence star because \citet{MarPomDehEA10} constrain its age to be $<3$ Myr.
The dynamical age of the shell is $\sim0.2$ Myr \citep{ArtHenMelEA11} assuming that a single massive star has formed in situ from a molecular cloud with mean density $n_0\approx1000\,\mathrm{cm}^{-3}$ (or somewhat older if the density is larger).
There are no other known nearby O stars but a number of young stellar objects have been discovered, particularly in the swept-up shell around the H\,\textsc{ii} region \citep{ZavPomDehEA07, DehZavSchEA09},  but see also \citet{WalWhiBisEA11}.

Pressure and density gradients in the ISM generate asymmetric H\,\textsc{ii} regions \citep[\emph{Champagne flows},][]{Ten79}, and this model was recently applied to RCW\,120 by \citet{OchVerCoxEA14} to explain the asymmetry implied by the mid-IR arcs.
Stellar motion also provides a pressure asymmetry that produces asymmetric stellar wind bubbles \citep{WeaMcCCasEA77, MacVanWooEA91, ArtHoa06} and H\,\textsc{ii} region bubbles \citep{Rag86, MacLanGva13}.
Massive stars are born in motion with respect to their surroundings because star formation is a dynamic process in which supersonic turbulence plays an important role \citep{McKOst07}.
Simulations show that the stars which form typically have small space velocities, $v_\star\approx 2-5\,\mathrm{km}\,\mathrm{s}^{-1}$ \citep[e.g.,][]{PetBanKleEA10,DalBon11}.
Furthermore, massive stars generally form in clusters \citep{LadLad03} and are very likely to be in binary or multiple systems \citep{SanDeKDeMEA13}.
Because of this, they often obtain more substantial space velocities through dynamical encounters with other nearby stars and from the disruption of binary systems when the primary star explodes \citep[see discussion in ][]{HooDeBDeZ01}.
\citet{EldLanTou11} predict that 20 per cent of all Type IIP supernova progenitors are runaway stars produced through the binary supernova scenario, and estimate that a similar percentage arise from runaways produced by dynamical ejections from star clusters.

It is therefore important to study the feedback from massive stars that are moving with various velocities through their surroundings.
H\,\textsc{ii} regions have typical temperatures $T_\mathrm{i}\approx6\,000-10\,000$ K and isothermal sound speeds $a_\mathrm{i}\approx10\,\mathrm{km}\,\mathrm{s}^{-1}$.
For velocities $v_\star>a_\mathrm{i}$ the supersonic wind-ISM interaction forms a bow shock \citep{BarKraKul71}, and for $v_\star>2a_\mathrm{i}$ a complete shell cannot form around the H\,\textsc{ii} region because the upstream ionization front becomes R-type \citep{Kah54}.
These two characteristic velocities divide the parameter space into three regimes, or four regimes if we consider $v_\star=0$ as a special case.
The cases $v_\star=0$ and $v_\star\geq 2a_\mathrm{i}$ have been well-studied but less work has been done on the two intermediate regimes that are the focus of this work.

Stellar wind bubbles \citep{WeaMcCCasEA77, GarLanMac96, GarMacLan96} and H\,\textsc{ii} regions \citep{MelArtHenEA06, KruStoGar07, WhaNor08, ArtHenMelEA11} around static stars have been investigated in detail with numerical simulations.
For an O7 star ($M_\star\approx30\,\msun$) embedded in an ISM with number density $n=1\,\mathrm{cm}^{-3}$, \citet{WeaMcCCasEA77} estimate that it would take up to 4.7 Myr before the wind bubble's shell can trap the H\,\textsc{ii} region.
Both \citet{WeaMcCCasEA77} and \citet{CapKoz01} derive criteria determining when a wind bubble can trap the H\,\textsc{ii} region, finding that strong winds (i.e.\ high mass stars) in dense gas are the most favourable scenario.
This prediction was verified by radiation-hydrodynamics simulations of both wind and radiative feedback for static O stars \citep{FreHenYor03,FreHenYor06,ToaArt11}.
Only for a $60\,\msun$ star did the wind almost-completely fill its H\,\textsc{ii} region; for $35-40\,\msun$ stars the wind bubble remained a distinct structure within the H\,\textsc{ii} region.
We therefore expect that wind bubbles are contained within a small to moderate fraction of the H\,\textsc{ii} region volume, at least for stars with mass $M_\star\lesssim40\,\msun$ in the first few Myr of their lives.
If the ISM is clumpy, then the evolution of the H\,\textsc{ii} region and wind bubble is more complicated \citep{McKVanLaz84, DalNgoErcEA14}, because the photoevaporating clumps act as a source of mass within the bubbles.

Many authors have modelled bow shocks from runaway O and B stars \citep[e.g.,][]{MacVanWooEA91, ComKap98, MeyMacLanEA14}, and there have been a few studies of H\,\textsc{ii} regions for $v_\star\geq20\,\mathrm{km}\,\mathrm{s}^{-1}$ \citep{TenYorBod79, RagNorCanEA97, MacLanGva13}.
In both cases a partial shocked shell forms, upstream for the bow shock and in the lateral direction for the H\,\textsc{ii} region, and it can be unstable in certain circumstances.

% What we do here that is new
In this work we explore the parameter space for stars moving with $4\,\mathrm{km}\,\mathrm{s}^{-1} \leq v_\star \leq 16\,\mathrm{km}\,\mathrm{s}^{-1}$, i.e.\ excluding the static case and the case where a complete shell cannot form.
In previous work, \citet{TenYorBod79} showed that a complete but asymmetric shell can form around the H\,\textsc{ii} region.
\citet{FraGarKurEA07} found strong ionization-front instabilities break up the H\,\textsc{ii} region shell for stars moving with these velocities from a dense to a less dense medium.
Neither of these studies included the stellar wind, and our aim here is to study simultaneously how both the wind bubble and H\,\textsc{ii} region respond to stellar motion and to each other.
We choose parameters for the star and ISM such that we can approximately compare our results with observations of the young H\,\textsc{ii} region RCW\,120 (see below), but the results are more generally applicable to stars moving slowly through dense gas.
Our study is similar to, and builds on, the work of \citet{ArtHoa06}, who modelled bow shocks and H\,\textsc{ii} regions around static and moving stars.
The difference is that we use significantly weaker winds, appropriate for stars with $M_\star\approx 30\,\msun$, so that the wind bubble remains significantly smaller than the H\,\textsc{ii} region.

The numerical methods and simulation setup are described in Sect.~\ref{sec:theory}.
Our results for four simulations are presented in Sect.~\ref{sec:results}.
We discuss our results in the context of previous work in Sect.~\ref{sec:discussion} and conclude in Sect.~\ref{sec:conclusions}.

\begin{figure}[h]
\centering
\resizebox{0.88\hsize}{!}{\includegraphics{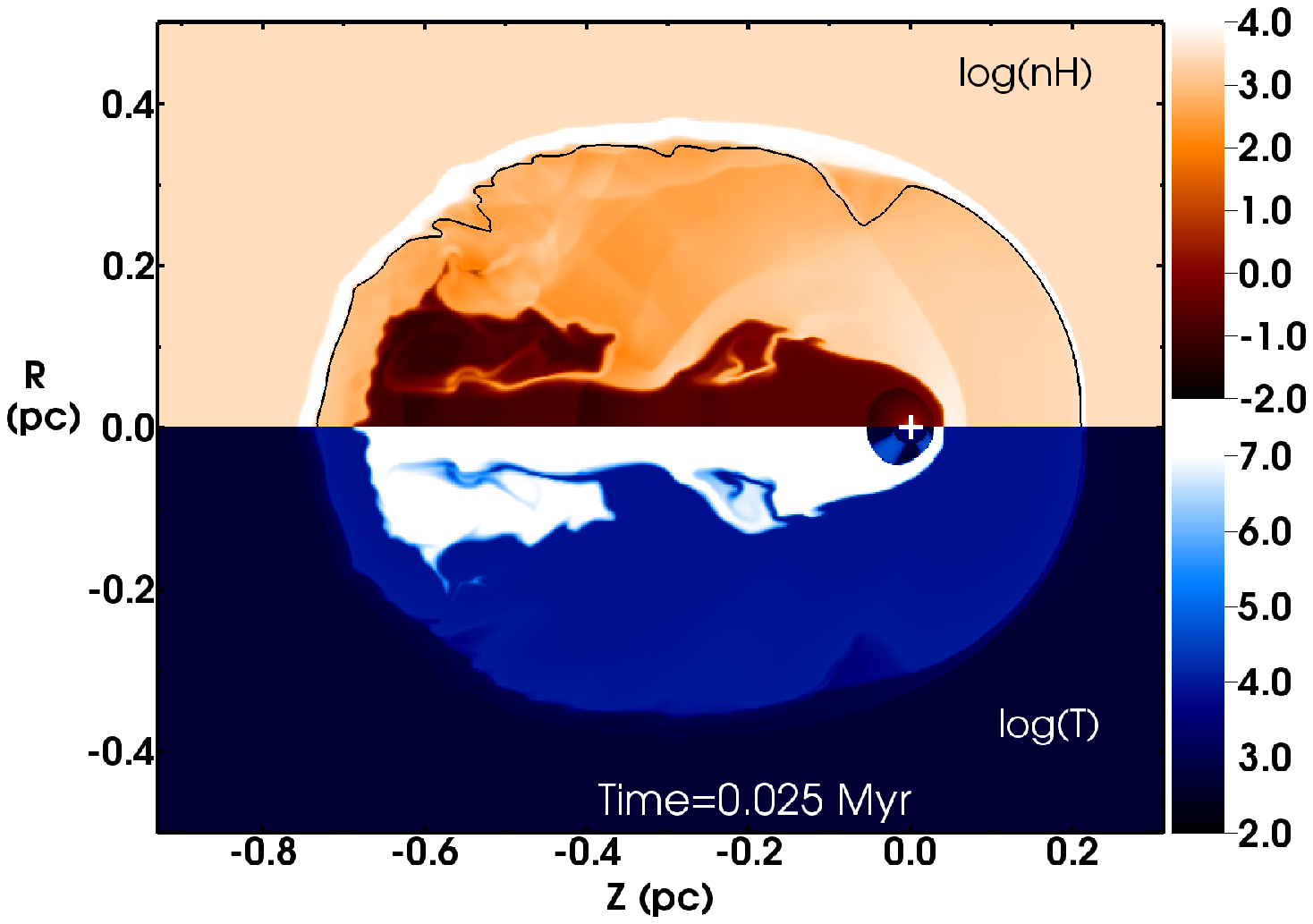}}
\resizebox{0.88\hsize}{!}{\includegraphics{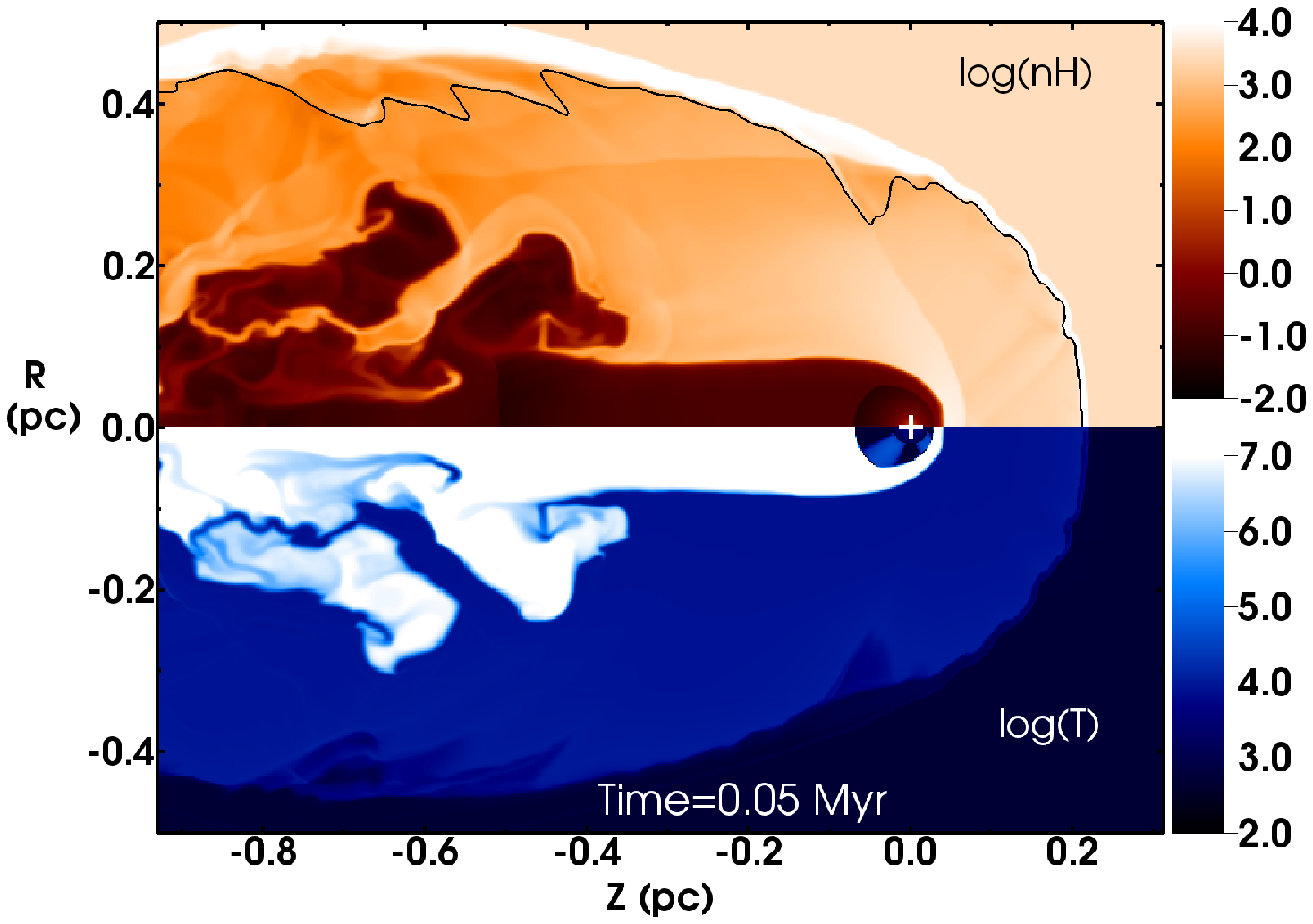}}
\resizebox{0.88\hsize}{!}{\includegraphics{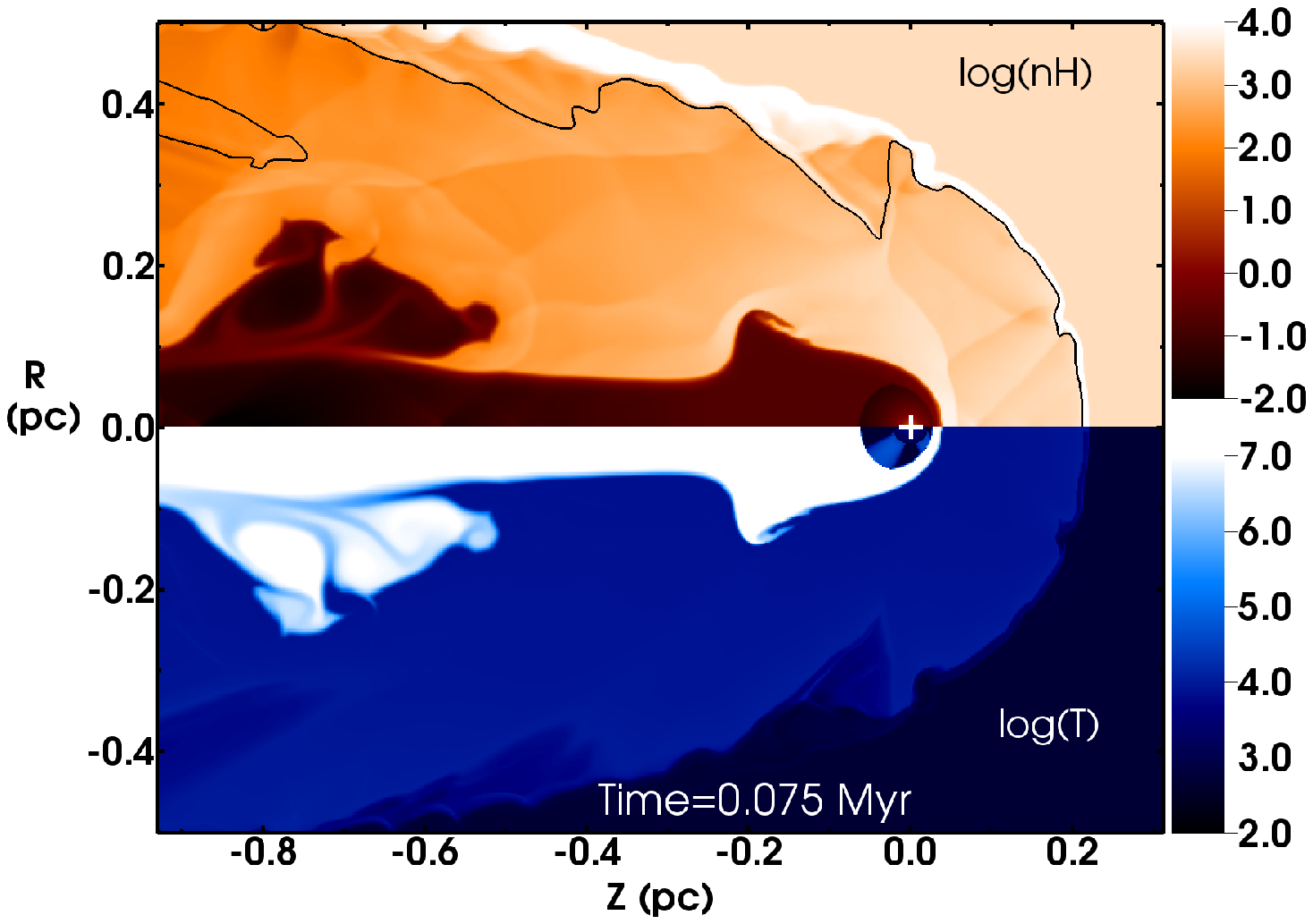}}
\caption{
  Log of gas number density (upper half-plane) and temperature (lower half-plane) for simulation V16 of a star moving with $v_\star=16\,\mathrm{km}\,\mathrm{s}^{-1}$ through a uniform ISM, with units $\log(n_{\mathrm{H}}{}/\mathrm{cm}^{-3})$ and $\log(T/\mathrm{K})$, respectively.
  The radial direction is vertical, and the axis of symmetry is $R=0$, the $z$-axis.
  The solid black contour in the upper half-plane shows ionization fraction $x=0.5$.
  The three panels show the evolution at different times, 0.025, 0.050, and 0.075 million years (Myr), respectively.
  The star is at the origin, denoted by a white cross.
  An animation of the simulation's evolution is available online.
  }
\label{fig:v16DT}
\end{figure}

%%% ------------------------------------------------------------
%%% ------------------------------------------------------------
\section{Numerical methods and initial conditions} \label{sec:theory}
%%% ------------------------------------------------------------
%%% ------------------------------------------------------------

We consider a star that emits extreme ultraviolet (EUV, with photon energy $h\nu>13.6$\,eV) ionizing photons, far ultraviolet (FUV, $6\leq h\nu\leq13.6$\,eV) photoheating photons, and a spherically-symmetric stellar wind.
The star is fixed on the simulation domain at the origin, and the ISM flows past with a relative velocity $v_\star$.
The photoionized H\,\textsc{ii} region grows rapidly to its equilibrium size, the \citet{Str39} radius $R_\mathrm{St}$, and the fast stellar wind drives a wind-blown bubble within the H\,\textsc{ii} region.
The relative motion between the star and the ISM means that the external pressure is asymmetric, so both the wind bubble and the H\,\textsc{ii} region also become distorted over time.
To model this system numerically we need:
\begin{enumerate}
\item
  at least two spatial dimensions,
\item
  a robust hydrodynamics solver to handle strong shocks,
\item
  a ray-tracer to calculate radiative transfer and attenuation of EUV and FUV photons emitted by the star,
\item
  a solver for the rate equation of H ionization,
\item 
  and heating and cooling rates that reflect the different processes occurring in ionized and neutral gas phases.
\end{enumerate}
We use the radiation-magnetohydrodynamics (R-MHD) code \textsc{pion} \citep{MacLim10, Mac12} for the simulations presented here.
The Euler equations of hydrodynamics are solved in two dimensions with cylindrical coordinates $(z,R)$ (assuming rotational symmetry about the axis $R=0$) on a uniform rectilinear grid.
We use a finite volume, shock-capturing integration scheme with geometric source terms to account for rotational symmetry \citep{Fal91}.
Inter-cell fluxes are calculated with Flux-Vector-Splitting \citep{vanLeer82}, a very robust algorithm (albeit diffusive) that is useful for high Mach number shocks.
The H ionization fraction, $x$, is advected with the flow using a passive tracer.

\begin{figure}
\centering
\resizebox{0.88\hsize}{!}{\includegraphics{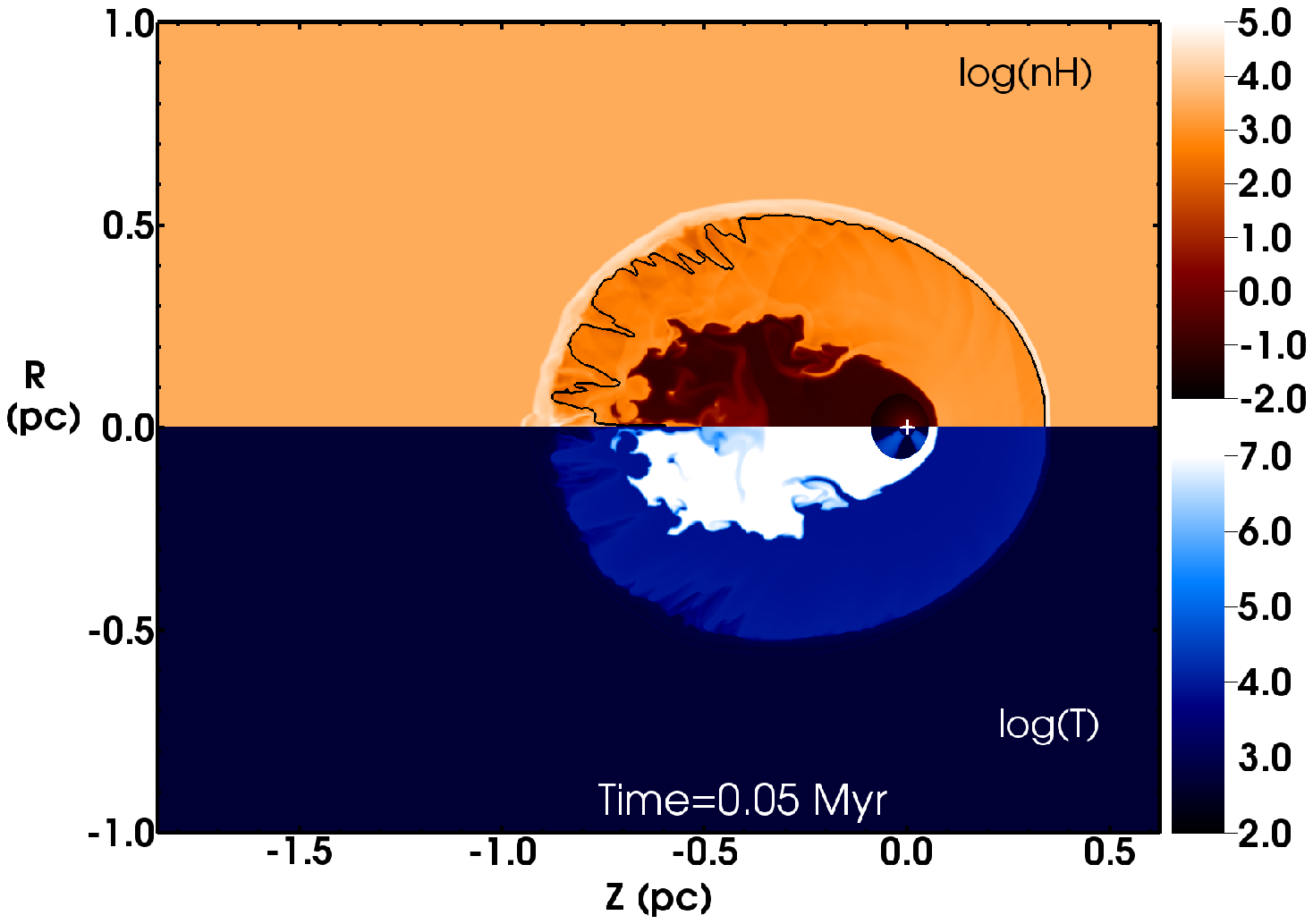}}
\resizebox{0.88\hsize}{!}{\includegraphics{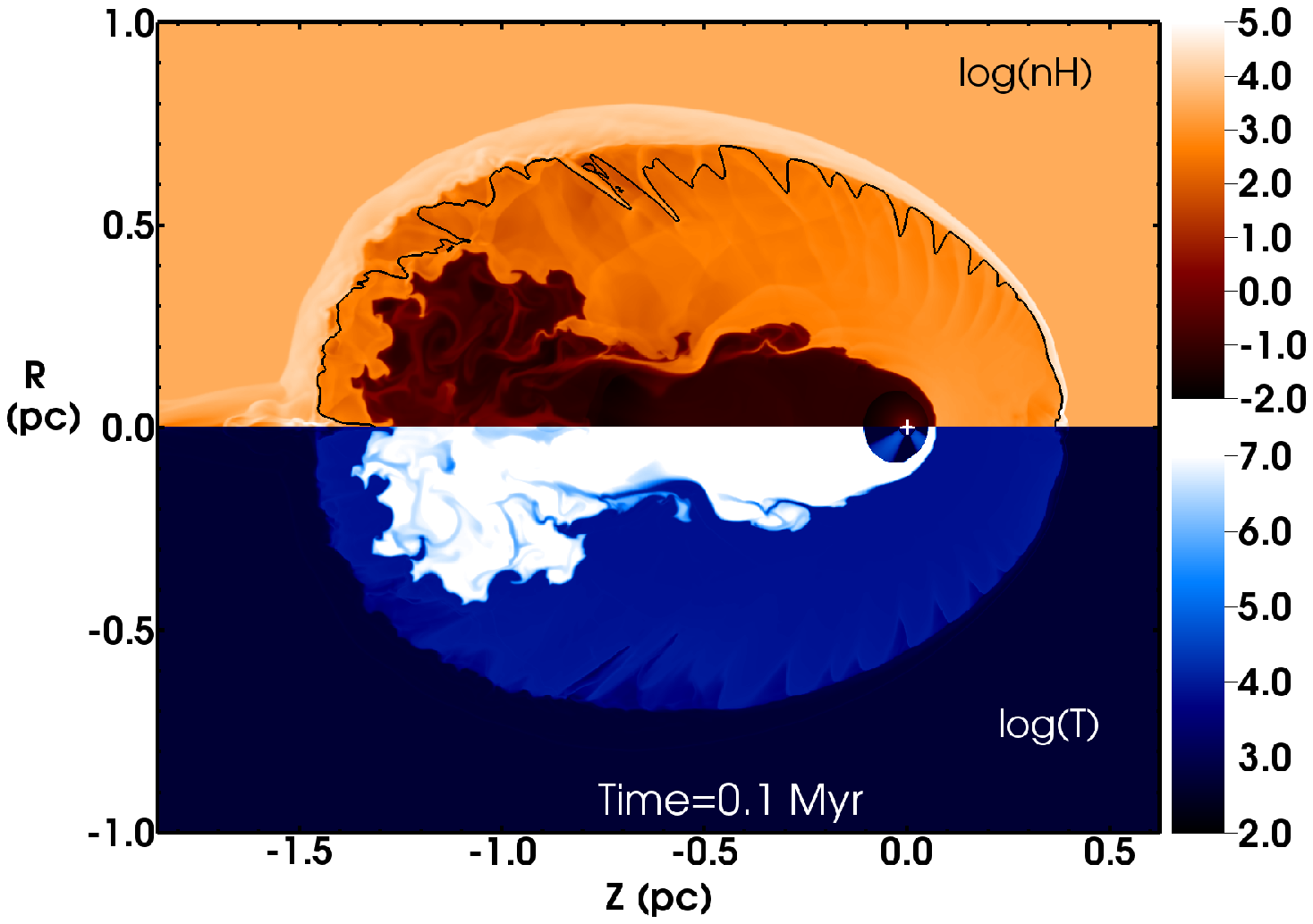}}
\resizebox{0.88\hsize}{!}{\includegraphics{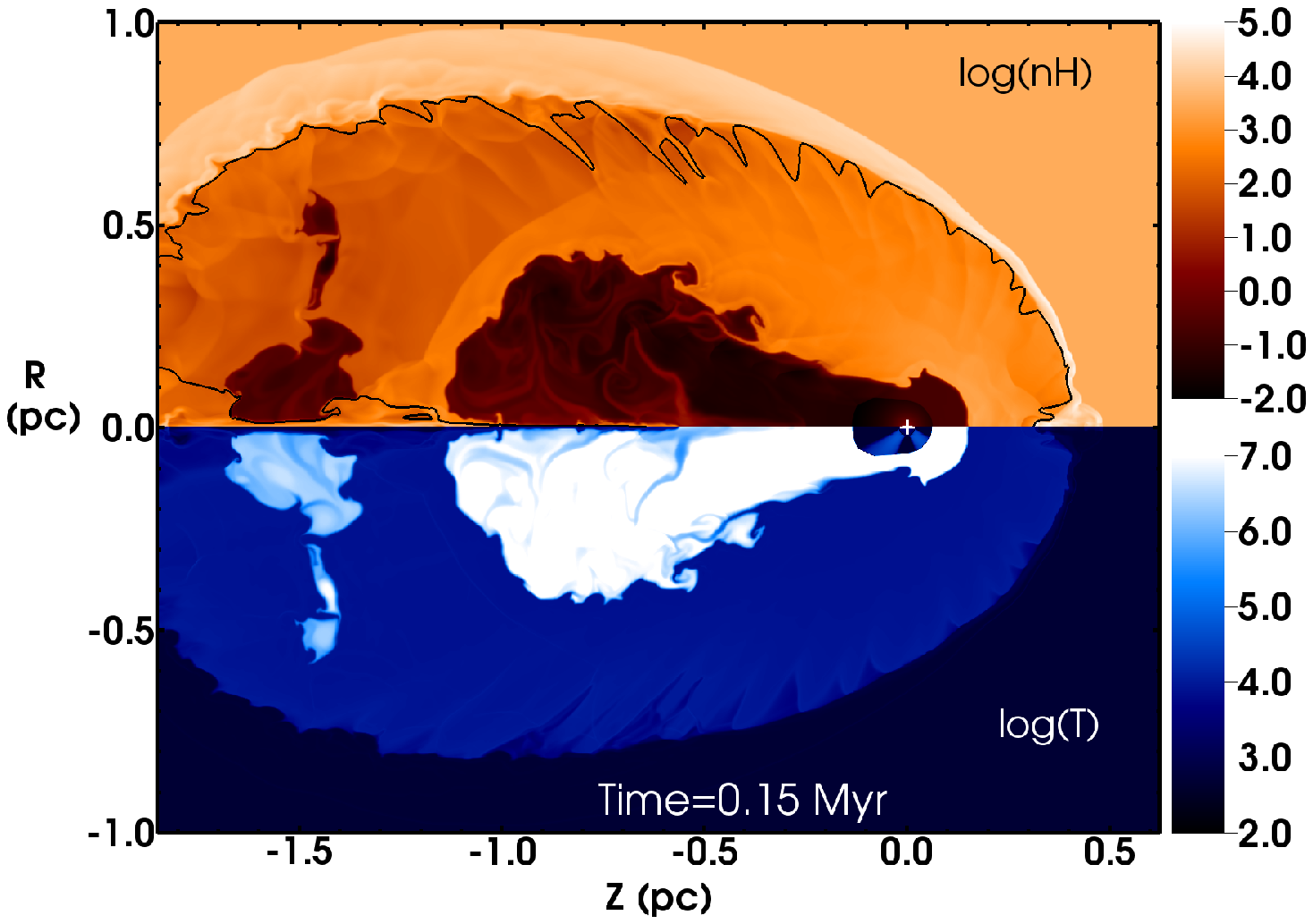}}
\caption{
  As Fig.~\ref{fig:v16DT}, but for simulation V08 of a star moving with $v_\star=8\,\mathrm{km}\,\mathrm{s}^{-1}$, and shown at different times.
  Now the star's motion is subsonic with respect to the H\,\textsc{ii} region, so there is no bow shock upstream from the star.
  An animation of the simulation's evolution is available online.
  }
\label{fig:v08DT}
\end{figure}

%%% ------------------------------------------------------------
\subsection{Ionization and thermal physics} \label{ssec:cooling}
%%% ------------------------------------------------------------
The H mass fraction is $X_\mathrm{H}=0.715$ \citep{AspGreSauEA09}, and we assume the rest is He when calculating the mean mass per particle.
The mean mass per H atom is $\mu_\mathrm{H}=m_\mathrm{p}/X_\mathrm{H}=1.399m_\mathrm{p}$ and the mean mass per nucleon is $\mu_\mathrm{n}=m_\mathrm{p}[X_\mathrm{H} +(1-X_\mathrm{H})/4]^{-1}=1.272m_\mathrm{p}$, where $m_\mathrm{p}$ is the proton mass.
The H number density is $n_\mathrm{H}=\rho/\mu_\mathrm{H}$, and the ratio of H to He number density is $n_\mathrm{H}/n_\mathrm{He}=10.0$.

For simplicity we consider that He is singly ionized whenever H is (i.e.\ we do not solve the He ionization rate equation), so the electron fraction is $n_\mathrm{e}/n_\mathrm{H} = 1.1x$.
This underestimates the electron fraction  by 10 per cent where He is doubly-ionized, but this is a small effect.
For these assumptions, the mean mass per particle in photoionized gas is $\mu=0.636 m_\mathrm{p}$.
The non-equilibrium, ionization rate equation for atomic H is solved including collisional ionization, radiative recombination, and photoionization.
We use a photon-conserving algorithm for multifrequency ionizing radiation transfer  \citep{MelIliAlvEA06} with a short characteristics ray-tracer to calculate attenuation of radiation between the source and a given grid zone \citep{RagMelArtEA99}.
We use the on-the-spot approximation which means we only consider the direct radiation from the star and not the scattered radiation.
Absorption of EUV photons by dust is not considered.

Gas heating and cooling is largely treated as in \citet{MacLanGva13}.
We use a collisional-ionization equilibrium cooling curve (including Bremsstrahlung) for high temperature gas with $T\gtrsim2\times10^{4}$\,K  \citep{WieSchSmi09}; forbidden-line metal cooling in photoionized gas with $5000\,\mathrm{K}\leq T\lesssim2\times10^{4}$\,K; and metal-line and molecular cooling in neutral gas.
For the neutral gas cooling we use equations A10 and A14 in \citet{HenArtDeCEA09} instead of equations C5-C8 in \citet{MacLanGva13} because they are more appropriate for dense molecular gas.
Photoheating from EUV and FUV photons from the star are included as in \citet{MacLanGva13}, following \citet{HenArtDeCEA09}.

The gas thermal evolution is integrated together with the ionization rate equation as coupled ordinary differential equations using the \textsc{cvode} package \citep{CohHin96}.
The resulting source terms in the energy and advection equations are added to the finite volume integration following \citet{FalKomJoa98}.

%%% ------------------------------------------------------------
\subsection{Stellar parameters}
%%% ------------------------------------------------------------
We base the stellar properties on  CD\,$-$38$\deg$11636, the ionizing star of RCW\,120, with luminosity, $\log (L_\star/\lsun)=5.07\pm0.21$ and effective temperature, $T_\mathrm{eff}=37\,500\pm2000$\,K \citep{MarPomDehEA10}.
Its EUV photon luminosity is $Q_{0}\approx 3\times10^{48}$ s$^{-1}$, which we adopt for our simulations.
We assume a blackbody spectrum, so $T_\mathrm{eff}$ and $Q_{0}$ fully determine the radiation from the star.
The FUV photon luminosity (which weakly heats the shell around the H\,\textsc{ii} region) is set to $Q_{\mathrm{FUV}}=7.5\times10^{48}$ s$^{-1}$, appropriate for the above blackbody emission parameters.

For the quoted stellar parameters we estimate a mass-loss rate of $\log [\dot{M}/(\msunperyr)]=-6.81^{+0.27}_{-0.25}$ based on the stellar evolution models of \citet{BroDeMCanEA11} and mass-loss prescriptions of \citet{VinDeKLam01}, so we adopt the central value, $\dot{M}=1.55\times10^{-7} \,\msunperyr$.
This is consistent with the observational constraint $\dot{M}\lesssim10^{-7}\,\msunperyr$ \citep{MarPomDehEA10}.
We assume a wind velocity of $v_\mathrm{w}=2000\,\mathrm{km}\,\mathrm{s}^{-1}$, although it has not been measured for this star \citep{MarPomDehEA10}.

The equilibrium temperature of the H\,\textsc{ii} region is $T_\mathrm{i}\approx7500$\,K for this radiation source and assuming solar metallicity ($T_\mathrm{i}$ increases near the H\,\textsc{ii} region's edge to $\approx9500$\,K because of spectral hardening).
The Str\"omgren radius around the star is $R_\mathrm{St}=(3Q_0/4\pi\alpha_\mathrm{B}n_\mathrm{e}n_\mathrm{H})^{1/3}$, where $\alpha_\mathrm{B}$ is the case B recombination coefficient of H \citep{Hum94} with a value $\approx3.3\times10^{-13}\,\mathrm{cm}^{3}\,\mathrm{s}^{-1}$ for $T=7500$\,K.
For the initial conditions we use here, we obtain $R_\mathrm{St}=0.20$\,pc.

%%% ------------------------------------------------------------
\subsection{Initial conditions and numerical details for the simulations}
%%% ------------------------------------------------------------

\begin{table}
  \centering
  \caption{
    Axisymmetric simulations that model the H\,\textsc{ii} region and wind bubble simultaneously.
    All simulations use the same wind properites and ionizing photon luminosity; see text for details.
    $v_\star$ is the stellar space velocity in $\mathrm{km}\,\mathrm{s}^{-1}$;
    $N_\mathrm{z}$ and $N_\mathrm{r}$ are the number of grid zones in the $\mathbf{\hat{z}}$ and $\mathbf{\hat{R}}$ directions, respectively;
    $[z]$ and $[R]$ are the simulation domain sizes in pc;
    $\tau_\mathrm{c}$ is the time it takes for the star to cross the simulation domain (most relevant timescales are substantially shorter than this); and
    $T_\mathrm{min}$ is the minimum temperature (in K) allowed in each simulation.
  }
  \begin{tabular}{ l c c c c c}
    %\hline
    ID  &  $v_*$ & $N_\mathrm{z}\times N_\mathrm{r}$ & ($[z]\times [R]$)  & $\tau_\mathrm{c}$ (Myr) & $T_\mathrm{min}$ (K)\\
    \hline
    V04 & 4 & $1280\times512$  & $5.0\times2.0$  & 1.22  & 300  \\
    V06 & 6 & $1280\times512$  & $3.75\times1.5$ & 0.611 & 500\\
    V08 & 8 & $1280\times512$  & $2.5\times1.0$  & 0.306 & 500 \\
    V16 & 16 & $640\times256$  & $1.25\times0.5$ & 0.076 & 500 \\
  \end{tabular}
  \label{tab:sims}
\end{table}

We have run a series of simulations that consider stars moving with $v_\star=4-16\,\mathrm{km}\,\mathrm{s}^{-1}$ through the ISM, described in Table~\ref{tab:sims}.
We use a uniform ISM with mass density, $\rho=7.021\times10^{-21}\,\mathrm{g\,cm}^{-3}$, corresponding to $n_\mathrm{H}=3000\,\mathrm{cm}^{-3}$.
The initial pressure is set to correspond to the minimum temperature allowed in the simulation, either 300 or 500 K (see below).
At these low velocities, the H\,\textsc{ii} region expansion has a significant effect on the photoionized gas properties and hence on the wind bubble, in contrast to simulations with $v_\star\gtrsim25\,\mathrm{km}\,\mathrm{s}^{-1}$ \citep{MacLanGva13}.

The separation of size-scales between the bow shock and H\,\textsc{ii} region is only a factor of a few, so it is possible to model the two simultaneously on a uniform grid.
The wind velocity is very large, however, so the timestep restrictions are severe and the computational requirements are significant, making 3D simulations prohibitively expensive.
Previous 3D simulations \citep{DalNgoErcEA14} avoid this restriction by simulating only the momentum input from the wind, and not the wind material itself.

Gas tends to pile up on the symmetry axis for axisymmetric simulations of unstable shells because of the coordinate singularity, so we have used two numerical techniques to counteract this.
For all simulations we set a minimum temperature, $T_\mathrm{min}$, for the shell (hence a maximum compression factor for a given $v_\star$), listed in Table~\ref{tab:sims}.
This was chosen by trial and error with low-resolution simulations, such that we use the lowest value of $T_\mathrm{min}$ possible which allows the simulation to run for at least half of a simulation crossing time, $\tau_\mathrm{c}$, before symmetry-axis artefacts start to dominate the solution.
This was sufficient for simulations V04 and V06, but not for the higher velocity simulations.
For these we take the more drastic step of switching off gas cooling within $3-5$ grid zones of $R=0$ in neutral gas.
The on-axis gas then becomes adiabatic, so any converging flows are reflected from $R=0$ without allowing gas to pile up on the axis.
This also seems to stabilise the H\,\textsc{ii} region shell to some extent, because the apex is naturally the point where instability would first appear.
This was necessary to allow the 2D simulations to evolve for a fraction of $\tau_\mathrm{c}$ before gas pileup on the symmetry axis destroys the solution; it would not be necessary in 3D simulations.

%%% ------------------------------------------------------------
%%% ------------------------------------------------------------
\section{Two-dimensional simulations with winds and ionization} \label{sec:results}
%%% ------------------------------------------------------------
%%% ------------------------------------------------------------

%%% ------------------------------------------------------------
\subsection{Supersonic simulation V16}
%%% ------------------------------------------------------------
Fig.~\ref{fig:v16DT} shows the evolution of simulation V16 at three times corresponding approximately to $0.33\tau_\mathrm{c},\,0.67\tau_\mathrm{c}$ and $\tau_\mathrm{c}$.
The almost-circular, low-density region around the star is the freely-expanding wind, and the hot, low-density region surrounding this and extending downstream (to the left) is shocked wind.
The postshock temperature in the wind bubble is $T_\mathrm{b}\approx6\times10^7$ K.
The shocked wind is separated from the photoionized ISM (the H\,\textsc{ii} region) by a strong contact discontinuity where mixing processes including Kelvin-Helmholz instability, but also numerical diffusion, are acting.
The photoionized ISM is separated from the neutral, undisturbed ISM, by a D-type ionization front and its associated shocked shell.
The supersonic motion of the star through the photoionized ISM generates a weak bow shock at $z\approx0.07$\,pc.

Most striking in Fig.~\ref{fig:v16DT} is how asymmetric the stellar wind bubble is, even at early times before the H\,\textsc{ii} region gets strongly distorted.
There is also strong turbulent mixing in the wake behind the star.
This is probably not captured very well with these simulations because the contact discontinuity is mediated by numerical diffusion in the numerical scheme rather than any physical process such as thermal conduction.
The degree of mixing will strongly affect the emission properties of the hot gas.

\begin{figure}
\centering
\resizebox{0.88\hsize}{!}{\includegraphics{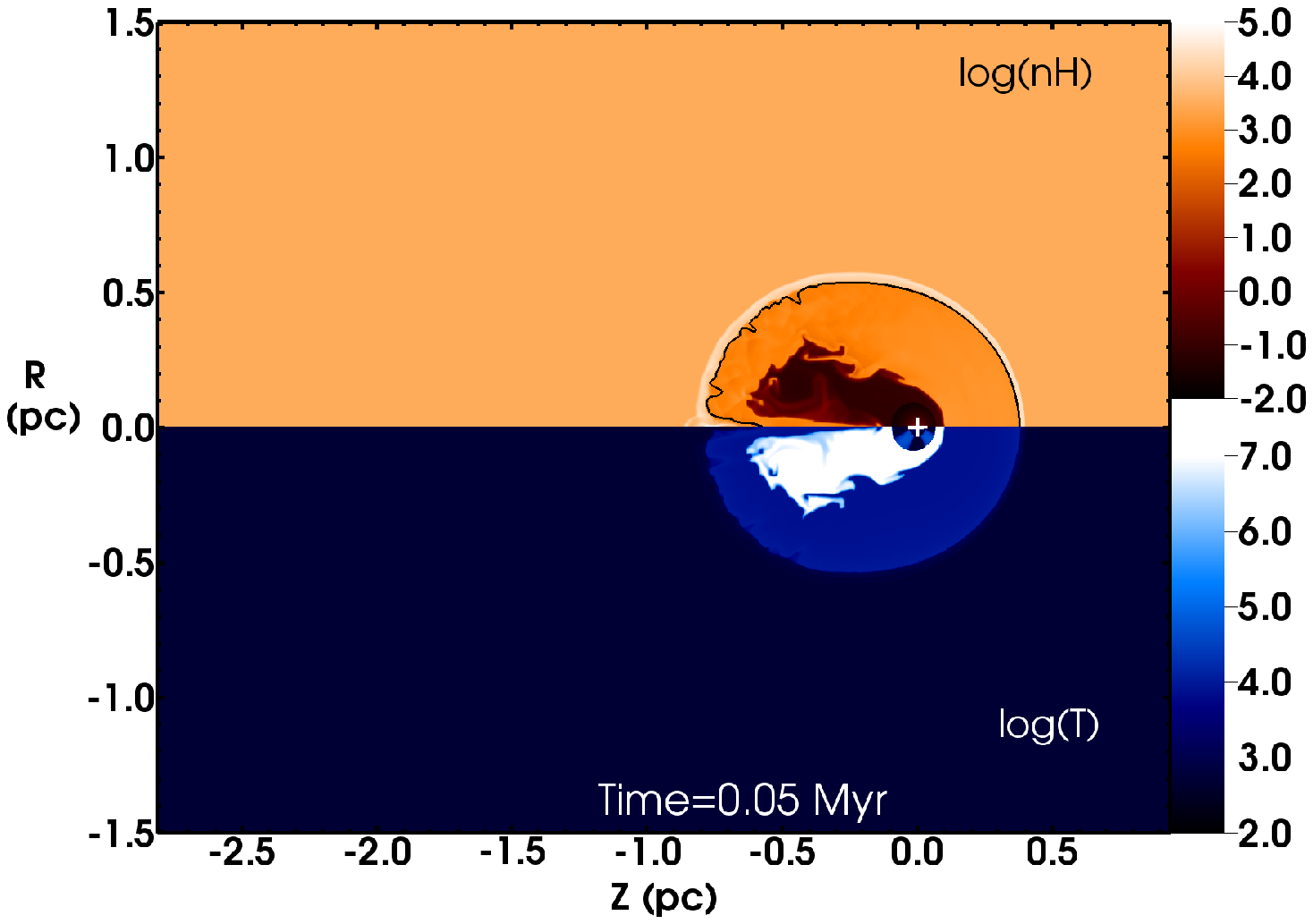}}
\resizebox{0.88\hsize}{!}{\includegraphics{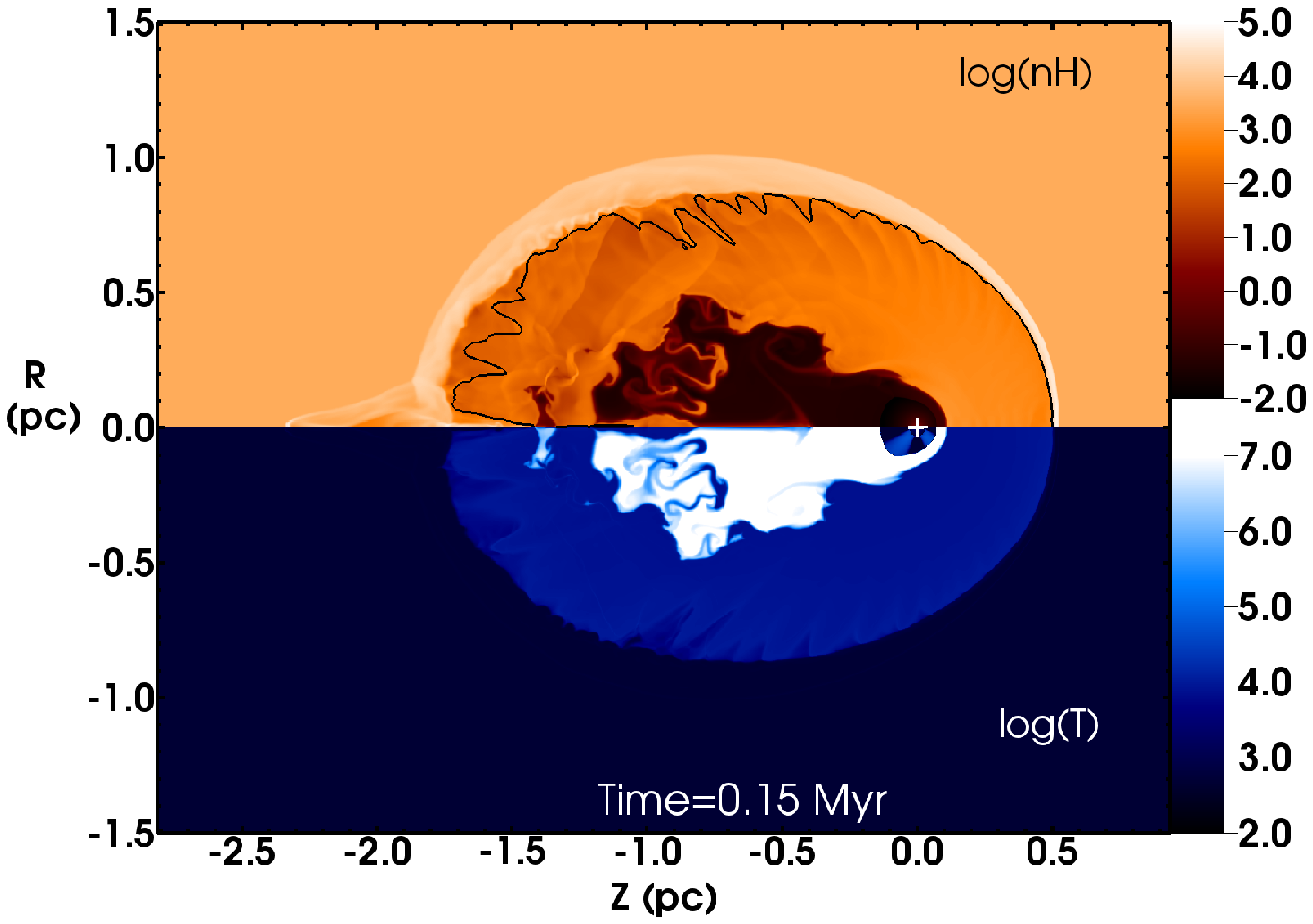}}
\resizebox{0.88\hsize}{!}{\includegraphics{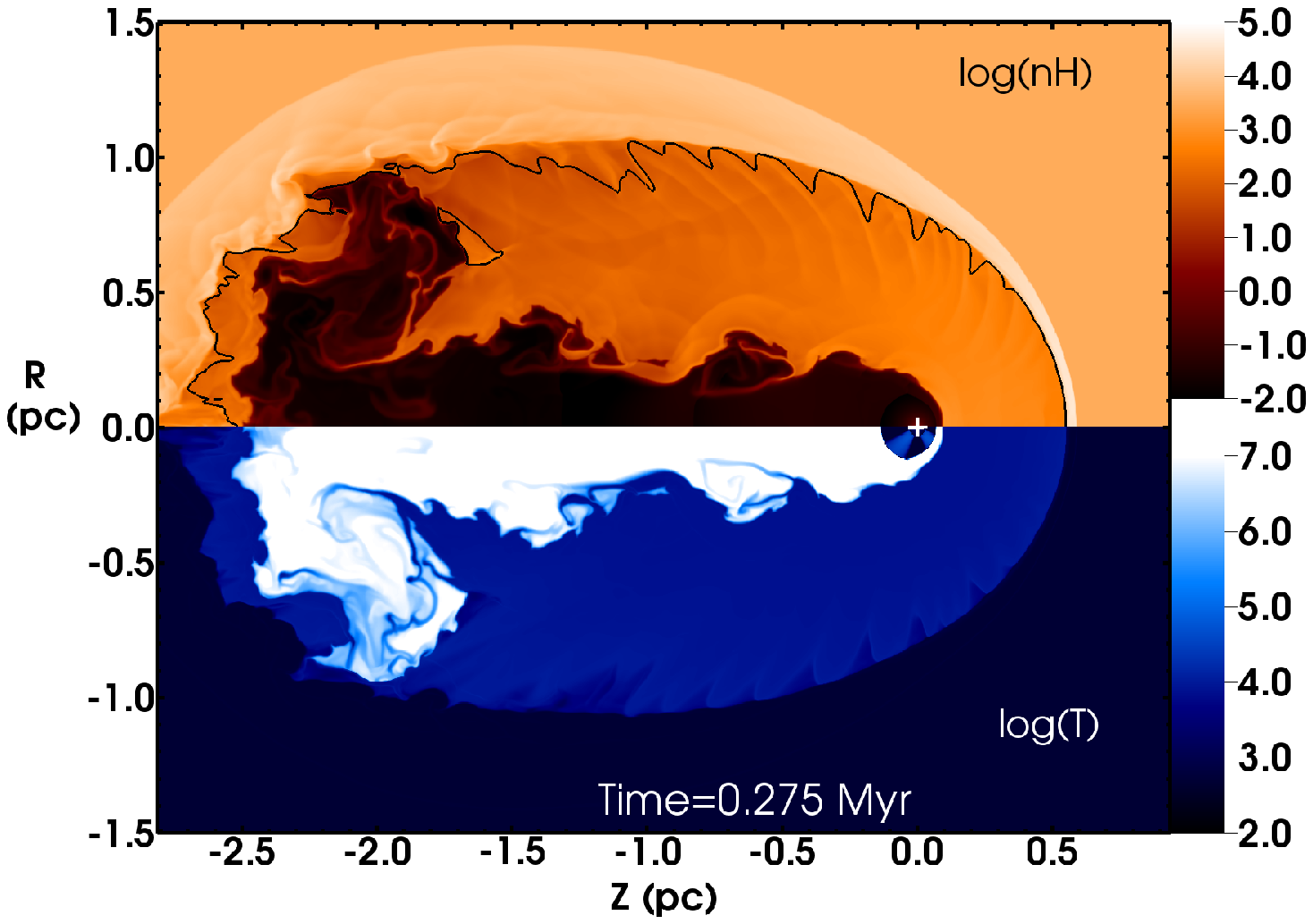}}
\caption{
  As Fig.~\ref{fig:v16DT}, but for simulation V06 of a star moving with $v_\star=6\,\mathrm{km}\,\mathrm{s}^{-1}$, and shown at different times.
  An animation of the simulation's evolution is available online.
  }
\label{fig:v06DT}
\end{figure}

\begin{figure*}
\centering
\resizebox{0.45\hsize}{!}{\includegraphics{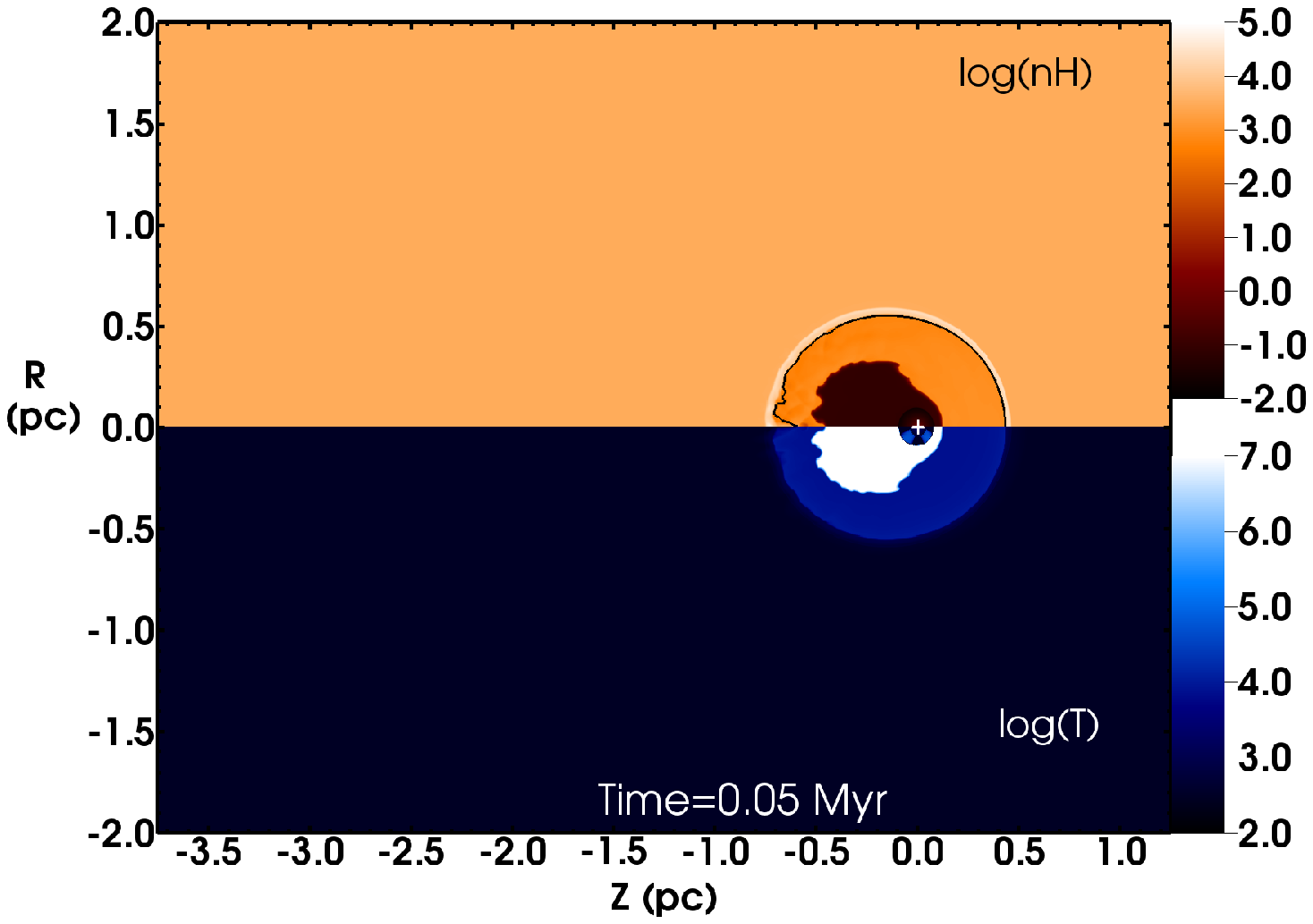}}
\resizebox{0.45\hsize}{!}{\includegraphics{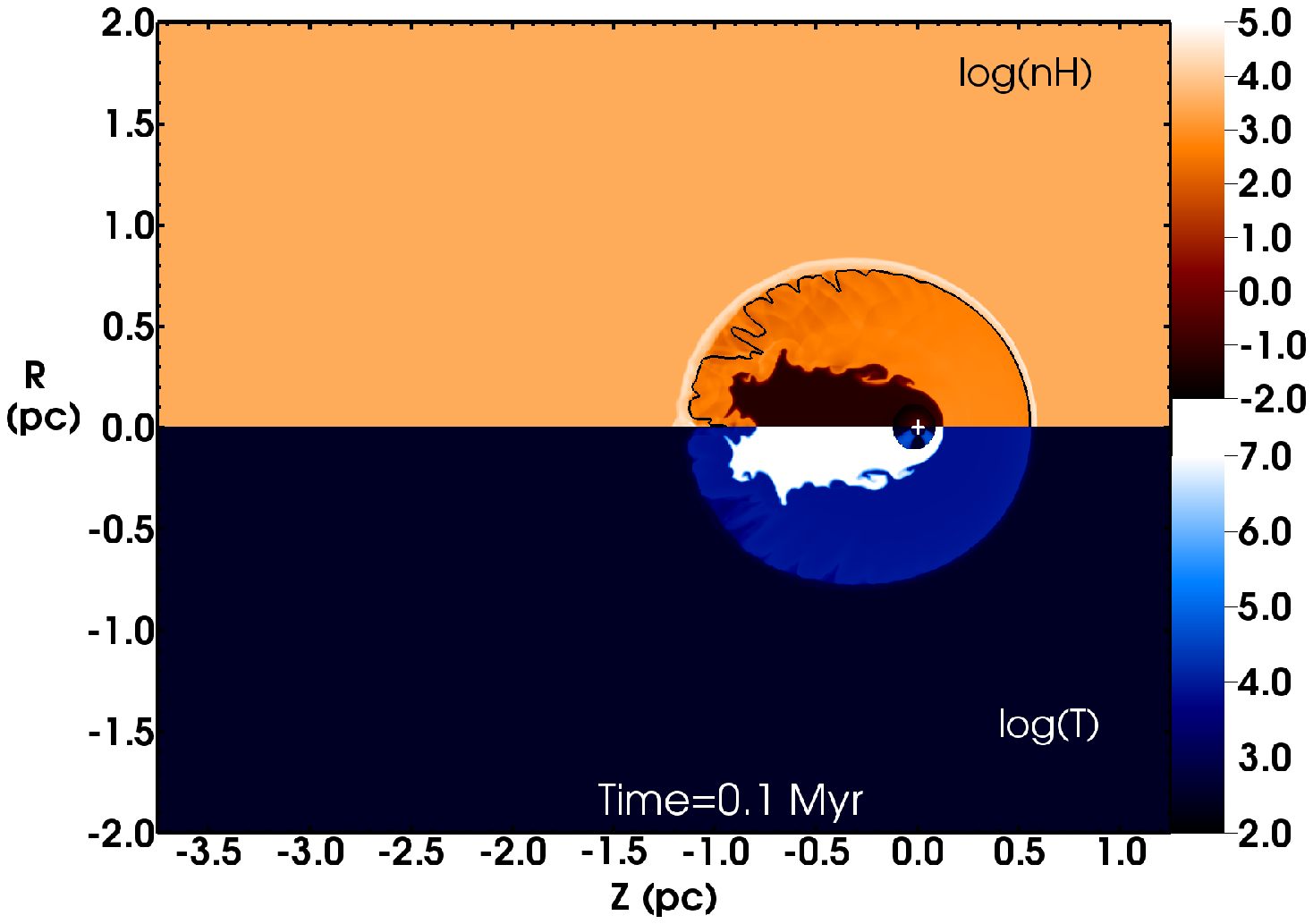}}
\resizebox{0.45\hsize}{!}{\includegraphics{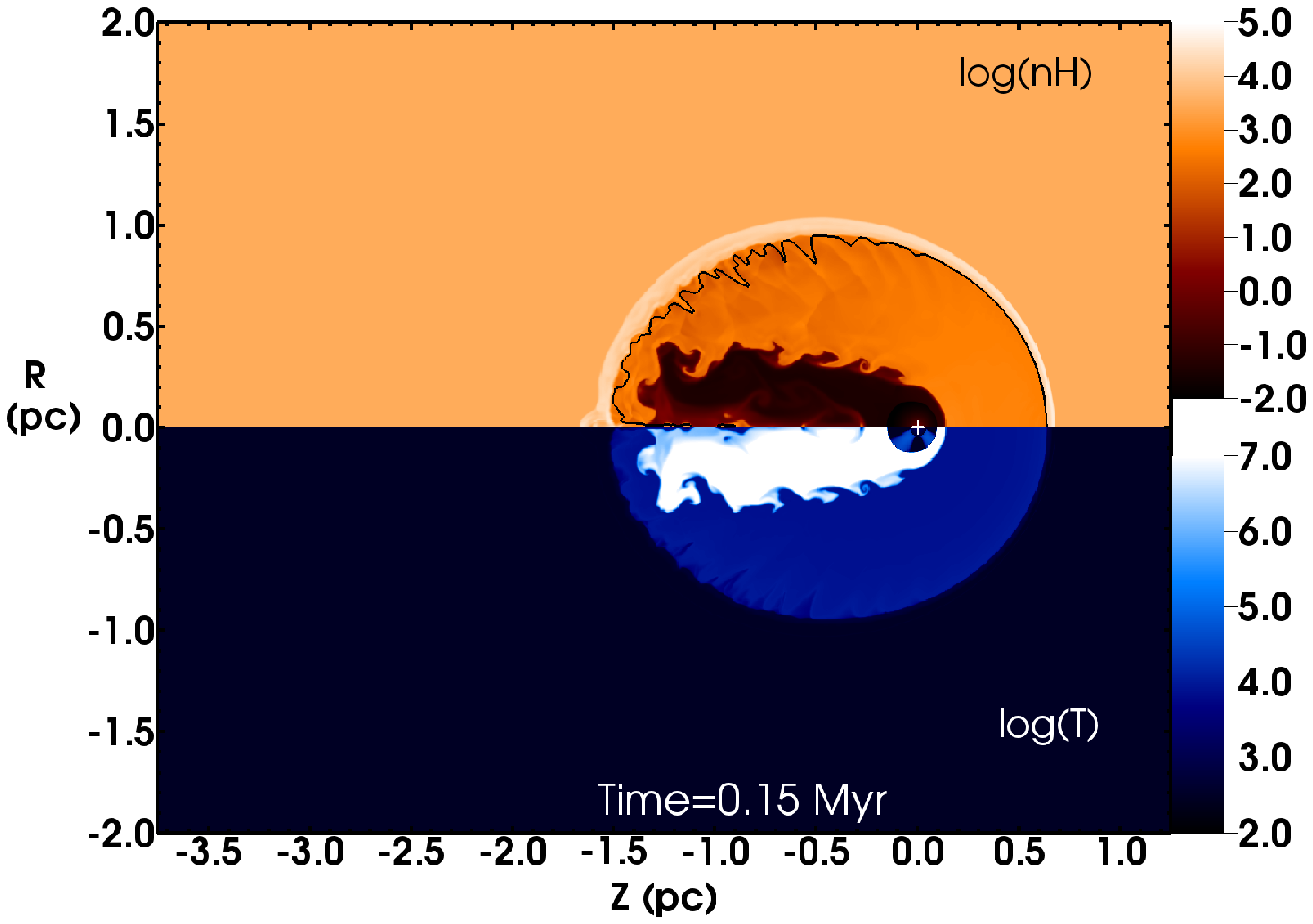}}
\resizebox{0.45\hsize}{!}{\includegraphics{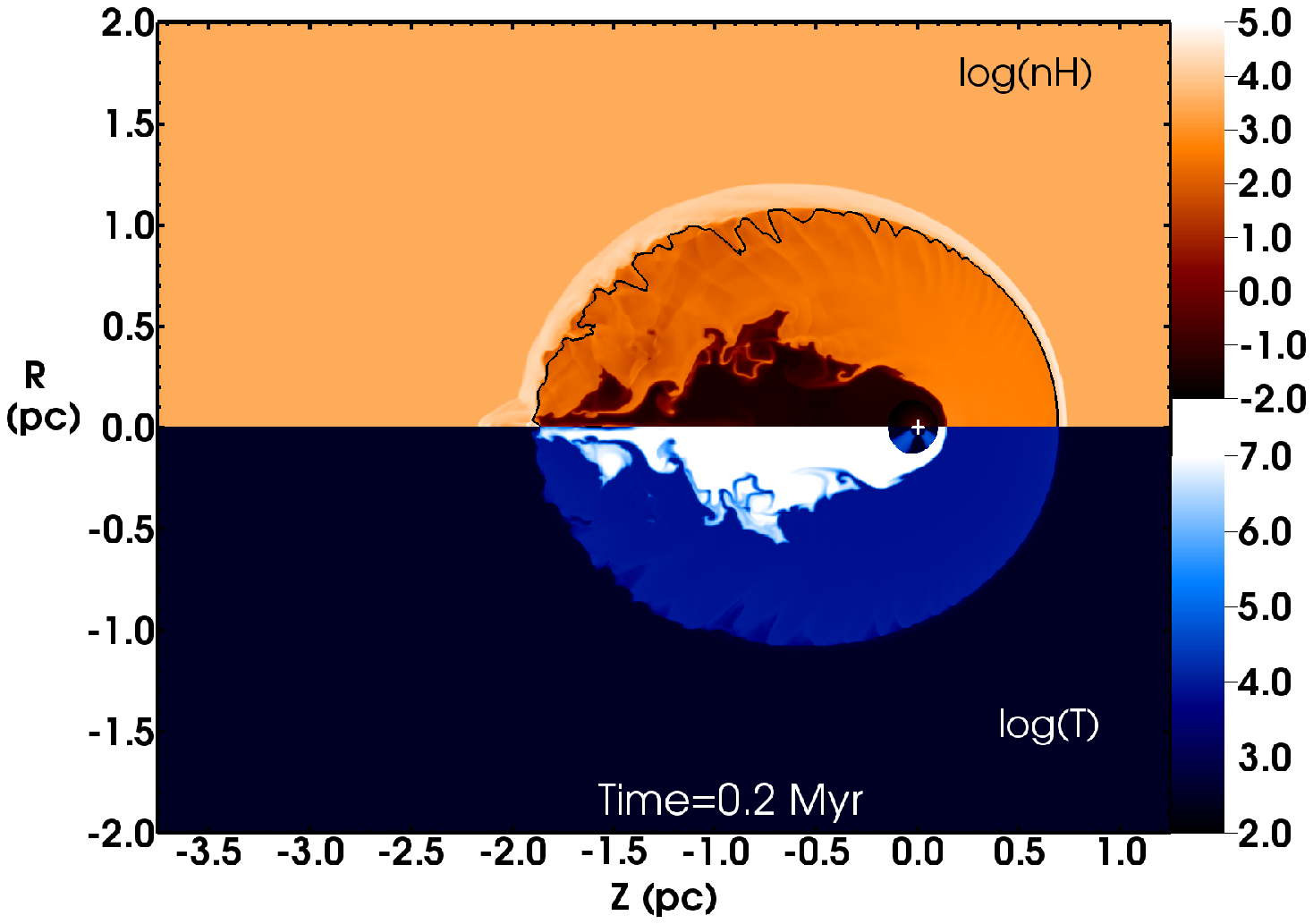}}
\resizebox{0.45\hsize}{!}{\includegraphics{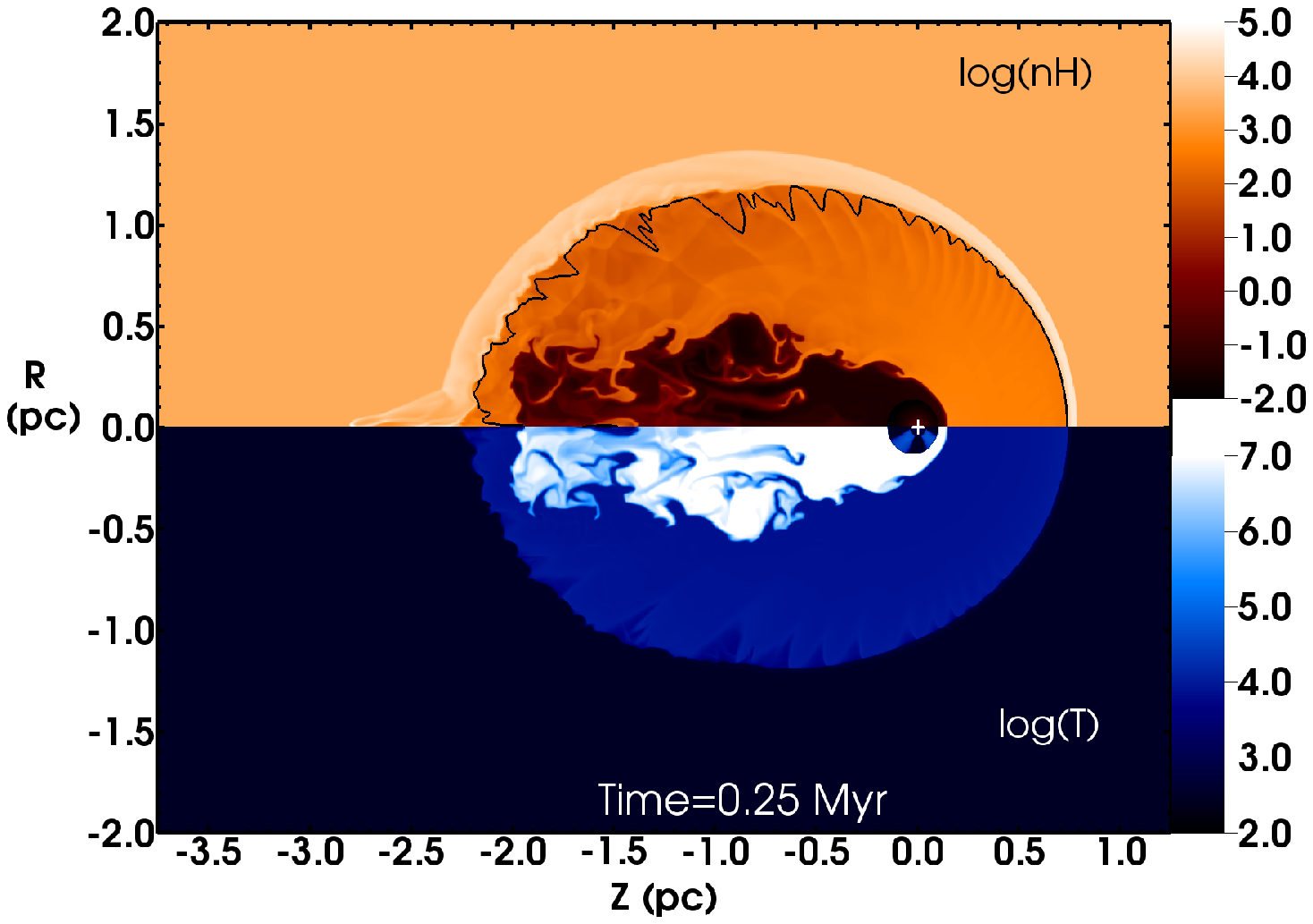}}
\resizebox{0.45\hsize}{!}{\includegraphics{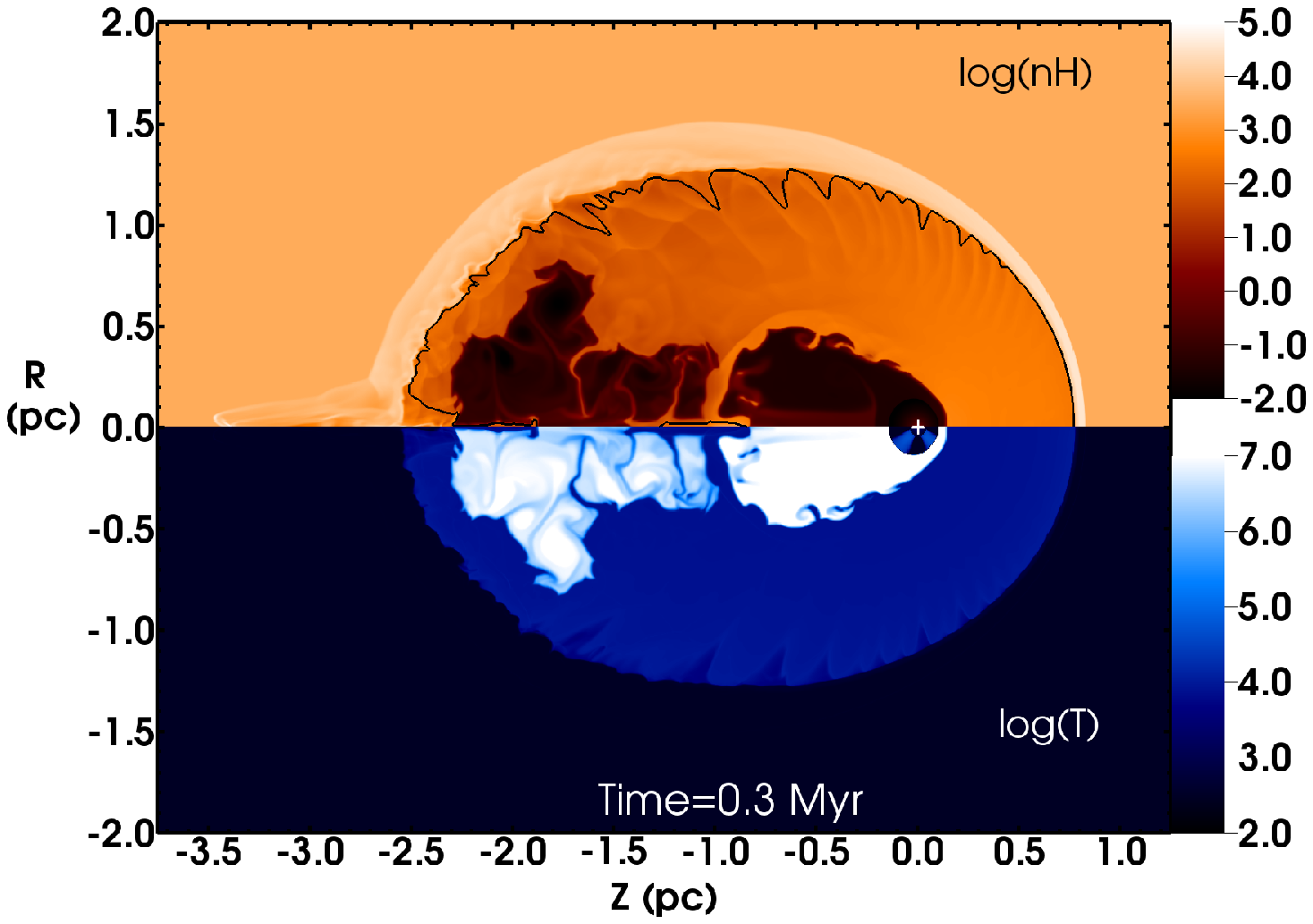}}
\resizebox{0.45\hsize}{!}{\includegraphics{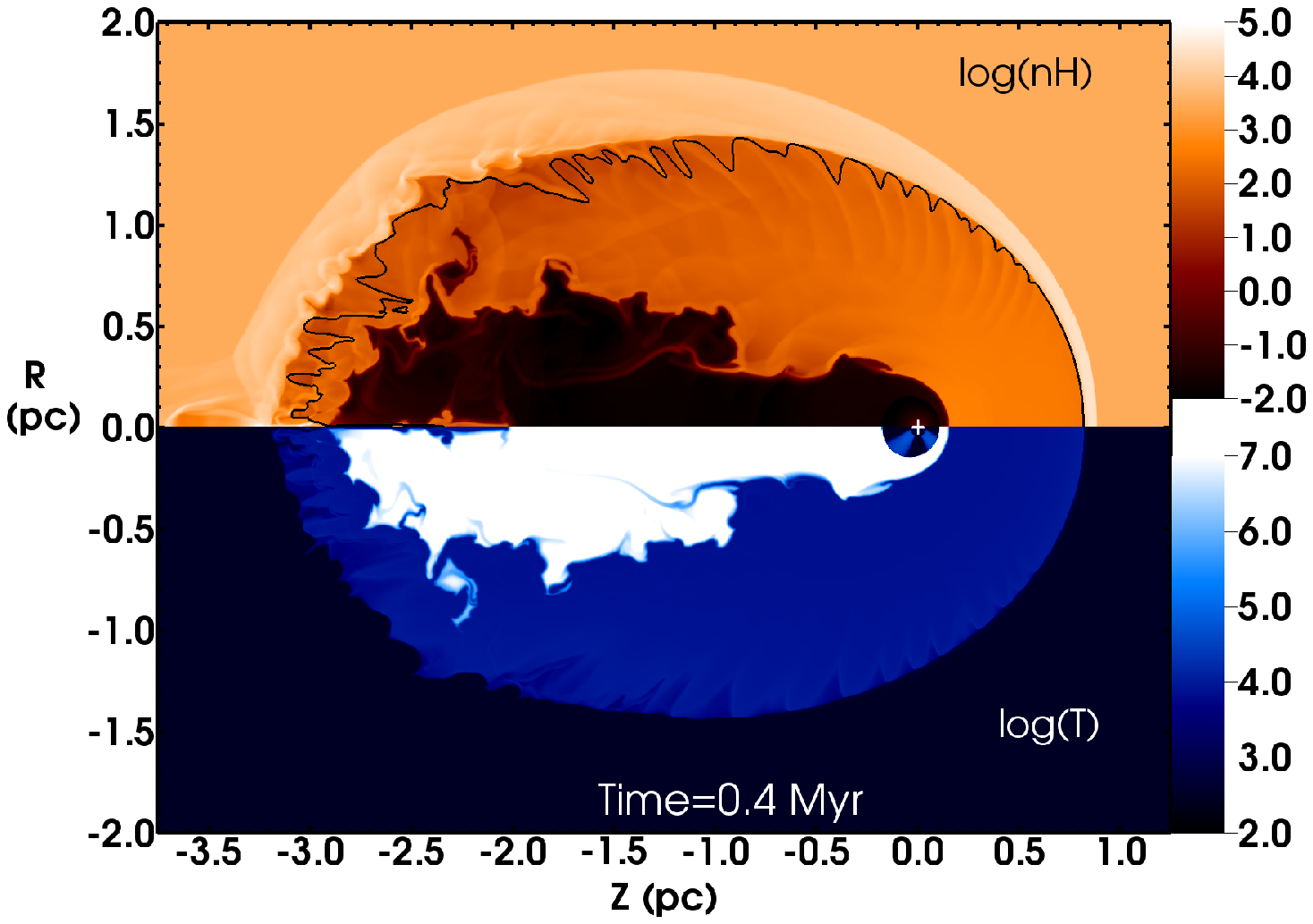}}
\resizebox{0.45\hsize}{!}{\includegraphics{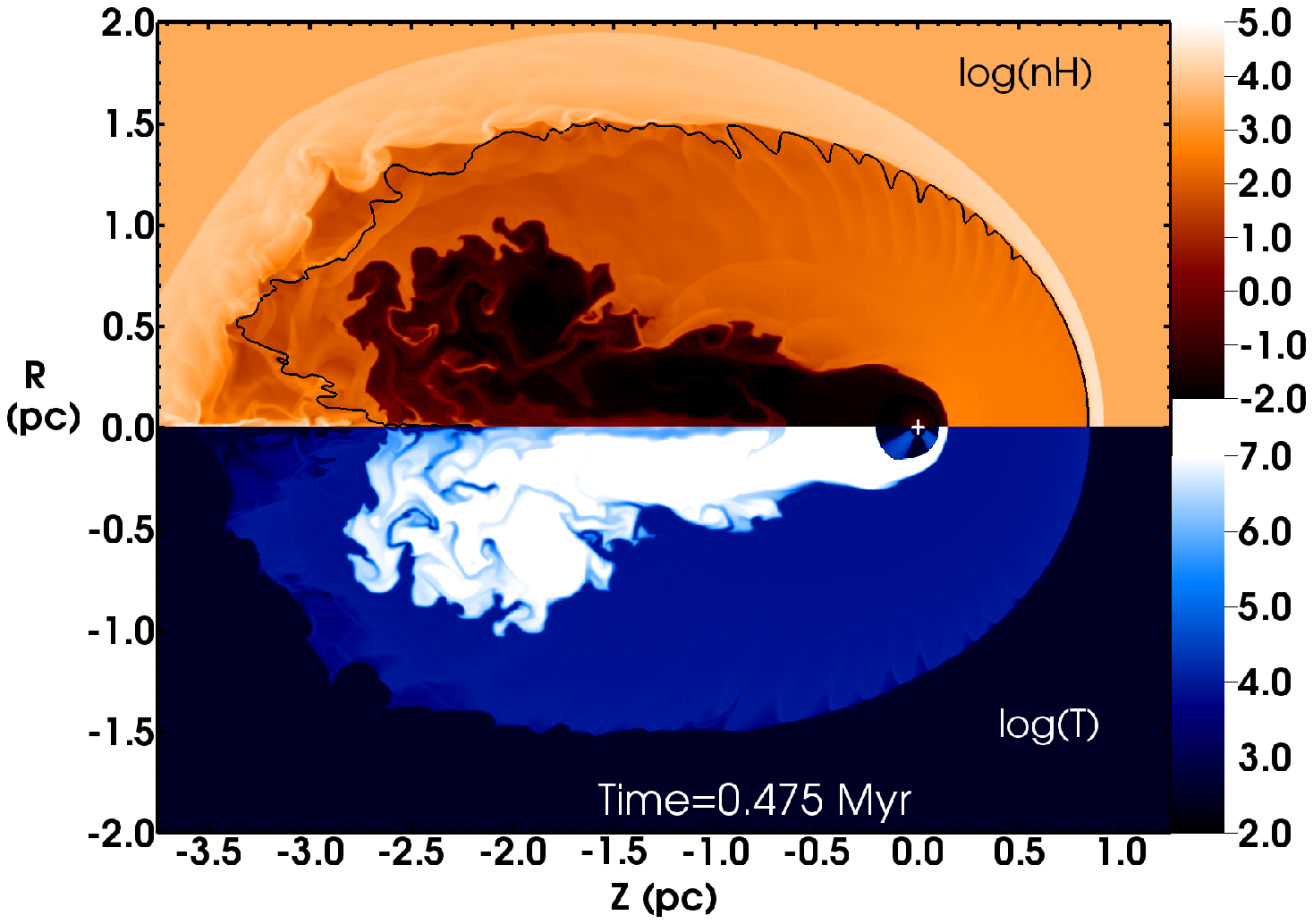}}
\caption{
  As Fig.~\ref{fig:v16DT}, but for simulation V04 of a star moving with $v_\star=4\,\mathrm{km}\,\mathrm{s}^{-1}$, and shown at different times.
  An animation of the simulation's evolution is available online.
  }
\label{fig:v04DT}
\end{figure*}

%%% ------------------------------------------------------------
\subsubsection{Contact discontinuity}
%%% ------------------------------------------------------------
The H\,\textsc{ii} region temperature $T_\mathrm{i}\approx7\,500$ K, so when the wind bubble and H\,\textsc{ii} region are in pressure equilibrium the density ratio between the shocked wind and photoionized ISM should be $\rho_\mathrm{i}/\rho_\mathrm{b}=T_\mathrm{b}/T_\mathrm{i}\approx8\,000$.
In the upstream direction the density in the bow shock is $n_{\mathrm{H}}\approx6000\,\mathrm{cm}^{-3}$ and in the upstream shocked wind bubble is $n_{\mathrm{H}}\approx0.7\,\mathrm{cm}^{-3}$, within 7 per cent of the theoretical expectation.
The huge temperature jump across this contact discontinuity should drive strong thermal conduction, which is not included here because of the extra computational expense in solving this parabolic equation.
\citet{ComKap98} and \citet{MeyMacLanEA14} showed that thermal conduction increases the size of a bow shock and reduces the temperature in most of the hot bubble, so it is likely that we are overestimating the hot bubble temperature and therefore somewhat underestimating its density.

On the other hand, observational determinations of the strength of thermal conduction in wind bubbles are few.
For the Wolf-Rayet bubble S308,  \citet{BorMcCClaEA97} measured blueshifted gas from the conduction front at the bubble border, and \citet{ChuGueGruEA03} estimate the thickness of this layer from the offset between the bubble boundary in X-ray and optical emission.
The conclusion on the strength of thermal conduction is not clear, and it may be that in many bubbles magnetic fields suppress the conduction \citep{CowMcK77}.
Given this uncertainty, it is still useful to run simulations without thermal conduction, to make testable predictions.

%%% ------------------------------------------------------------
\subsubsection{Bow shock}
%%% ------------------------------------------------------------
The H\,\textsc{ii} region number density is $n_{\mathrm{H}}\approx2400\,\mathrm{cm}^{-3}$, so the compression factor of the bow shock is only 2.5, corresponding to an isothermal Mach number $\mathcal{M}\equiv v_\mathrm{s}/a_\mathrm{i}=\sqrt{2.5}=1.58$, where $v_\mathrm{s}$ is the shock velocity and $a_\mathrm{i}\equiv\sqrt{p/\rho}=\sqrt{k_\mathrm{b}T/\mu m_\mathrm{p}}$ is the isothermal sound speed in the photoionized gas, with a value $a_\mathrm{i}=9.87\,\mathrm{km}\,\mathrm{s}^{-1}$ for $T_\mathrm{i}=7\,500$ K.
The shock is approximately isothermal because of the large gas density and consequent short cooling time.
If we equate $v_\star$ with $v_\mathrm{s}$, then we expect $\mathcal{M}=1.62$, again very close to what is measured in the simulations.
The shock velocity is actually  $v_\mathrm{s}\approx15\,\mathrm{km}\,\mathrm{s}^{-1}$, because the ISM gas is decelerated in the H\,\textsc{ii} region shell, and subsequently re-accelerated as it emerges from the D-type ionization front and moves down the pressure gradient within the asymmetric H\,\textsc{ii} region.
This is basically the same as the acceleration that occurs in a Champagne flow \citep{Ten79} from an H\,\textsc{ii} region that is only confined on one side, as studied in detail by \citet{ArtHoa06}, although the H\,\textsc{ii} region shape in a Champagne flow model strongly depends on the density stratification \citep[see e.g.,][]{OchVerCoxEA14}.

%%% ------------------------------------------------------------
\subsubsection{Comparison of H\,\textsc{ii} region size with bow shock size}
%%% ------------------------------------------------------------
The upstream radius of the H\,\textsc{ii} region, $R_\mathrm{up}$, is 0.21 pc, very similar to $R_\mathrm{St}=0.20$ pc.
This is because $v_\star$ is approaching $2a_\mathrm{i}\approx19.7\,\mathrm{km}\,\mathrm{s}^{-1}$, which is the maximum velocity of a D-type ionization front \citep{Kah54}.
In this limit, the shocked shell cannot propagate upstream beyond $R_\mathrm{St}$ because its maximum propagation velocity is equal to $v_\star$.
We do not expect exact agreement with theory because the H\,\textsc{ii} region shell traps much of the upstream ISM and so the upstream H\,\textsc{ii} region density is lower than the ISM density.
Furthermore, the bow shock is denser than the undisturbed ISM, even for this low Mach number.
Recombination rates are quadratic in gas density, so these under- and over-densities affect the actual value of $R_\mathrm{St}$ in any given direction.
In fact the overdense bow shock traps a sufficient number of photons that the H\,\textsc{ii} region develops a kink just downstream from perpendicular to the star's direction of motion, which could be observed in tracers of the ionized gas.

Apart from this kink perpendicular to the star's motion, the H\,\textsc{ii} region is \emph{not} trapped by the bow shock, in contrast to what some previous studies have assumed \citep{BreKah88,MacVanWooEA91} and simulated \citep{ArtHoa06}.
This difference arises because we use a weaker stellar wind than previous work.
We use $\dot{M}=1.55\times10^{-7}\,\msunperyr$, whereas the typical values considered by \citet{MacVanWooEA91} and \citet{ArtHoa06} are $\dot{M}\sim10^{-6}\,\msunperyr{}$.
If we increase $\dot{M}$ by $10\times$, then the bow shock will be $\sqrt{10}\times$ larger in radius, approximately coincident with the H\,\textsc{ii} region shell, so the two structures would almost certainly merge.

%%% ------------------------------------------------------------
\subsubsection{H\,\textsc{ii} region shell}
%%% ------------------------------------------------------------
The maximum density in the shell is in the upstream direction where the flow through the ionization front is fastest.
Making the approximation that the shocked shell is isothermal with this simulation's minimum temperature, $T_\mathrm{min}=500$ K, the Mach number of the shock is $\mathcal{M}=8.89$, so the shell number density should be $n_{\mathrm{H}}=3\,000\,\mathcal{M}^2 \,\mathrm{cm}^{-3}=2.4\times10^5\,\mathrm{cm}^{-3}$.
The shell density in the simulation increases from $3.7\times10^4\,\mathrm{cm}^{-3}$ at $t=0.025$ Myr to $2.0\times10^5\,\mathrm{cm}^{-3}$ at $t=0.075$ Myr, but at the latter time this maximum is only obtained in the dense knots that are forming in the shell.
The reason for this discrepency is that the shocked shell never cools below $T=2\,000$ K because of FUV heating (and limited spatial resolution), so the actual Mach number is lower than our estimate above.
The shell is not well resolved, in that the cooling length is comparable to the zone size, so it is likely that we are somewhat underestimating the shell density.

%%% ------------------------------------------------------------
\subsection{Subsonic simulations V08, V06, and V04}
%%% ------------------------------------------------------------

Results from simulation V08, plotted in Fig.~\ref{fig:v08DT}, show most of the same features as V16, with the notable exception that there is no stellar wind bow shock in the photoionized ISM because the star is now moving subsonically with respect to the  H\,\textsc{ii} region sound speed $a_\mathrm{i}$.
The stellar wind bubble is still very asymmetric, and the shocked wind material is rapidly swept downstream into a turbulent wake that piles up against (and reflects off) the H\,\textsc{ii} region shell.
Even at early times, after 0.050 Myr, the stellar wind bubble is almost entirely downstream.
By 0.150 Myr, dense gas has begun to accumulate on the symmetry axis upstream from the star, generating a strong photoevaporation flow which also distorts the wind bubble because the H\,\textsc{ii} region density and flow velocity are not constant in time.
This effect is significantly worse without the numerical fixes described in Section~\ref{sec:theory}, because the H\,\textsc{ii} region shell should be unstable.

Simulation V06, shown in Fig.~\ref{fig:v06DT}, is very similar to V08, except that the H\,\textsc{ii} region is a little less asymmetric.
Using $T_\mathrm{min}=500$ K for this simulation makes the H\,\textsc{ii} region shell sufficiently thick to be dynamically stable, but it should be borne in mind that by consequence we are significantly overestimating the shell thickness in all directions.
The turbulent wake is difficult to distinguish from that of V08, but both are rather different from V16 because there is no bow shock in the ISM.

The time evolution of simulation V04 is shown in Fig.~\ref{fig:v04DT}.
Even with the very low space velocity of $v_\star=4\,\mathrm{km}\,\mathrm{s}^{-1}$ the wind bubble is strongly asymmetric after 0.050 Myr, at which time the H\,\textsc{ii} region is still almost spherical.
The same turbulent wake develops as for V06 and V08, and for this case using $T_\mathrm{min}=300$ K was sufficient to stabilise the H\,\textsc{ii} region shell and prevent symmetry-axis artefacts from developing.
The H\,\textsc{ii} region for V04 is (as expected) the most spherical of all of the simulations, but the wind bubble again only fills the downstream part of the H\,\textsc{ii} region.

\begin{figure*}
\centering
\resizebox{0.45\hsize}{!}{\includegraphics{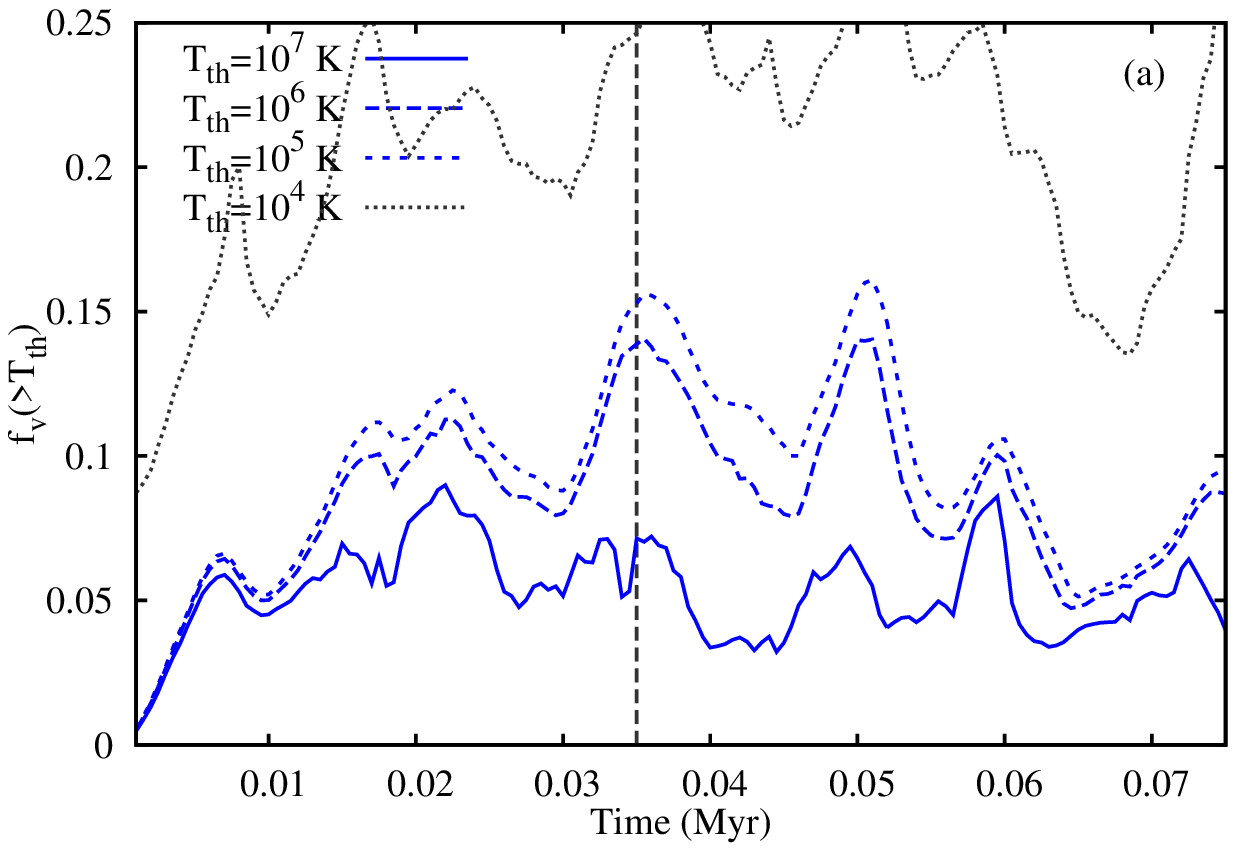}}
\resizebox{0.45\hsize}{!}{\includegraphics{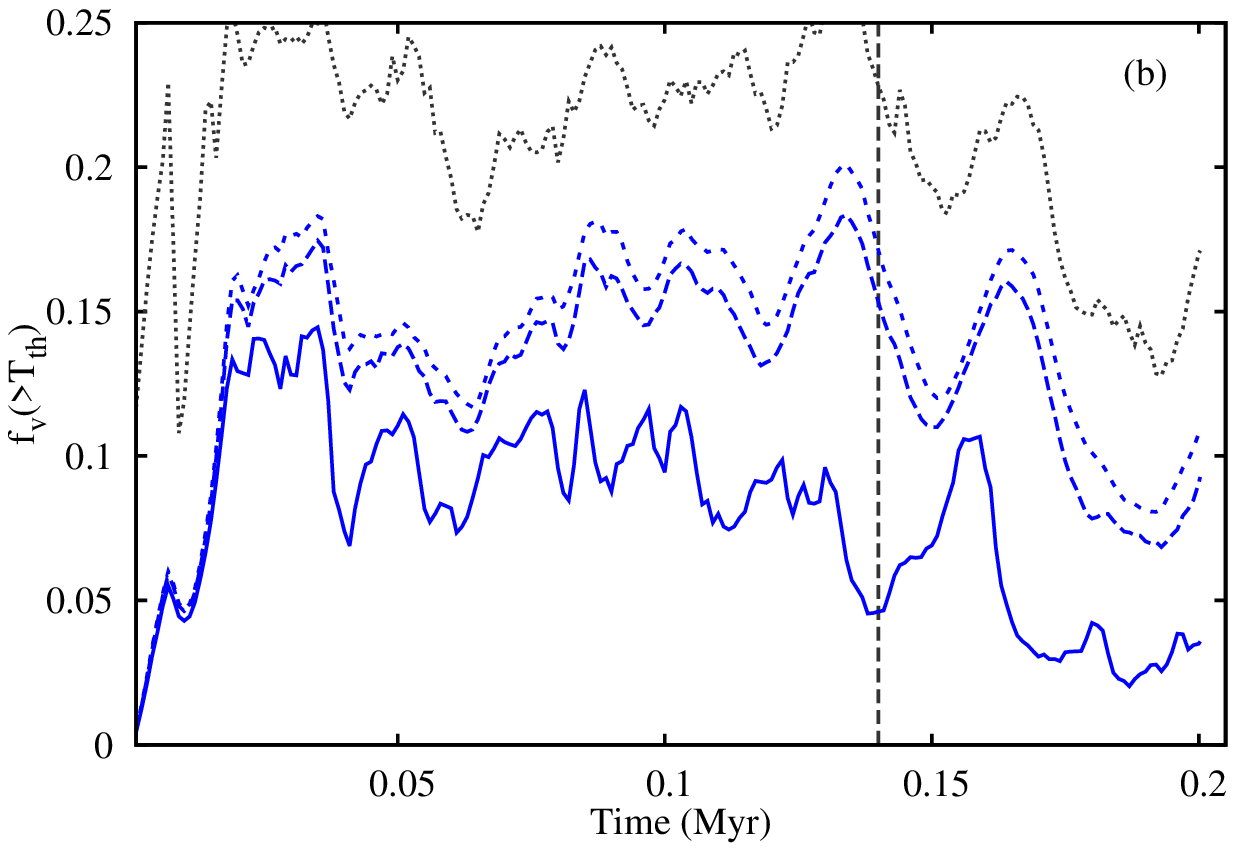}}
\resizebox{0.45\hsize}{!}{\includegraphics{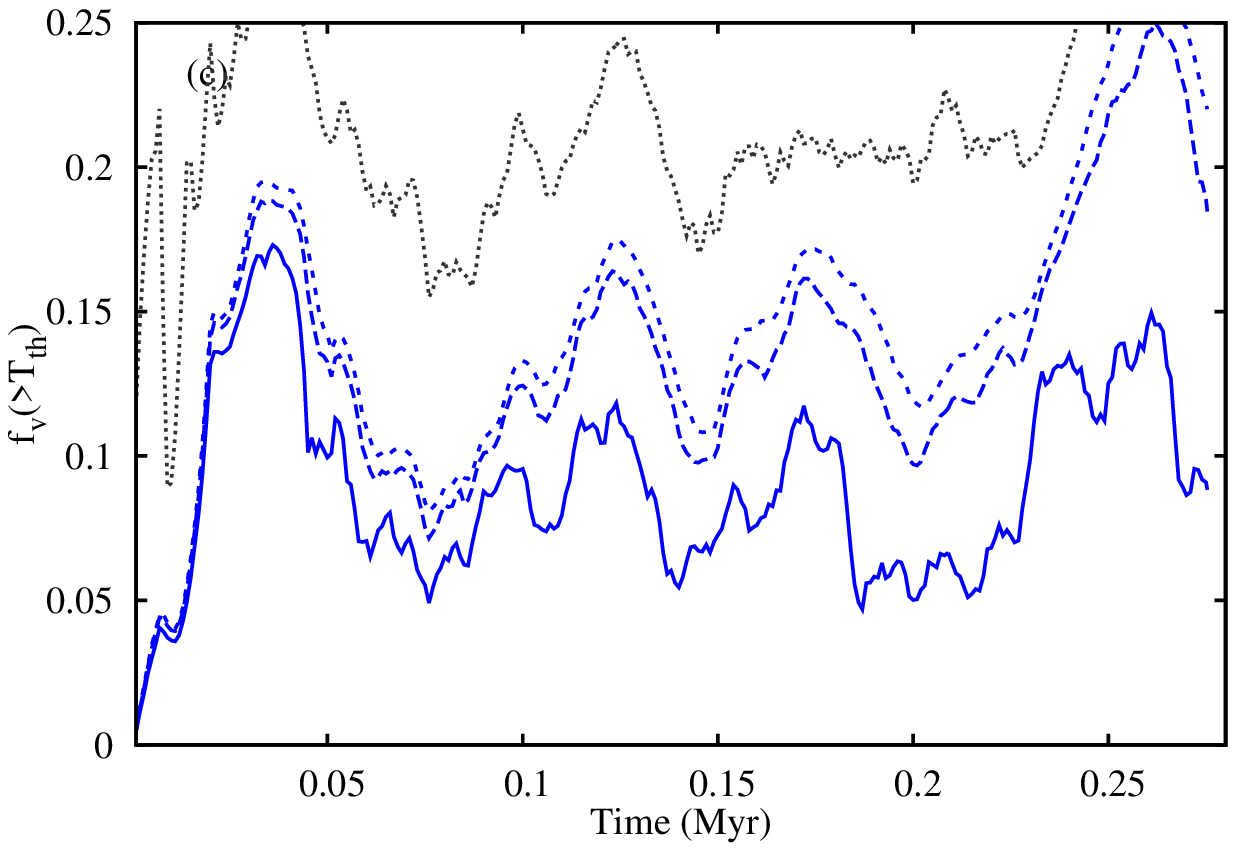}}
\resizebox{0.45\hsize}{!}{\includegraphics{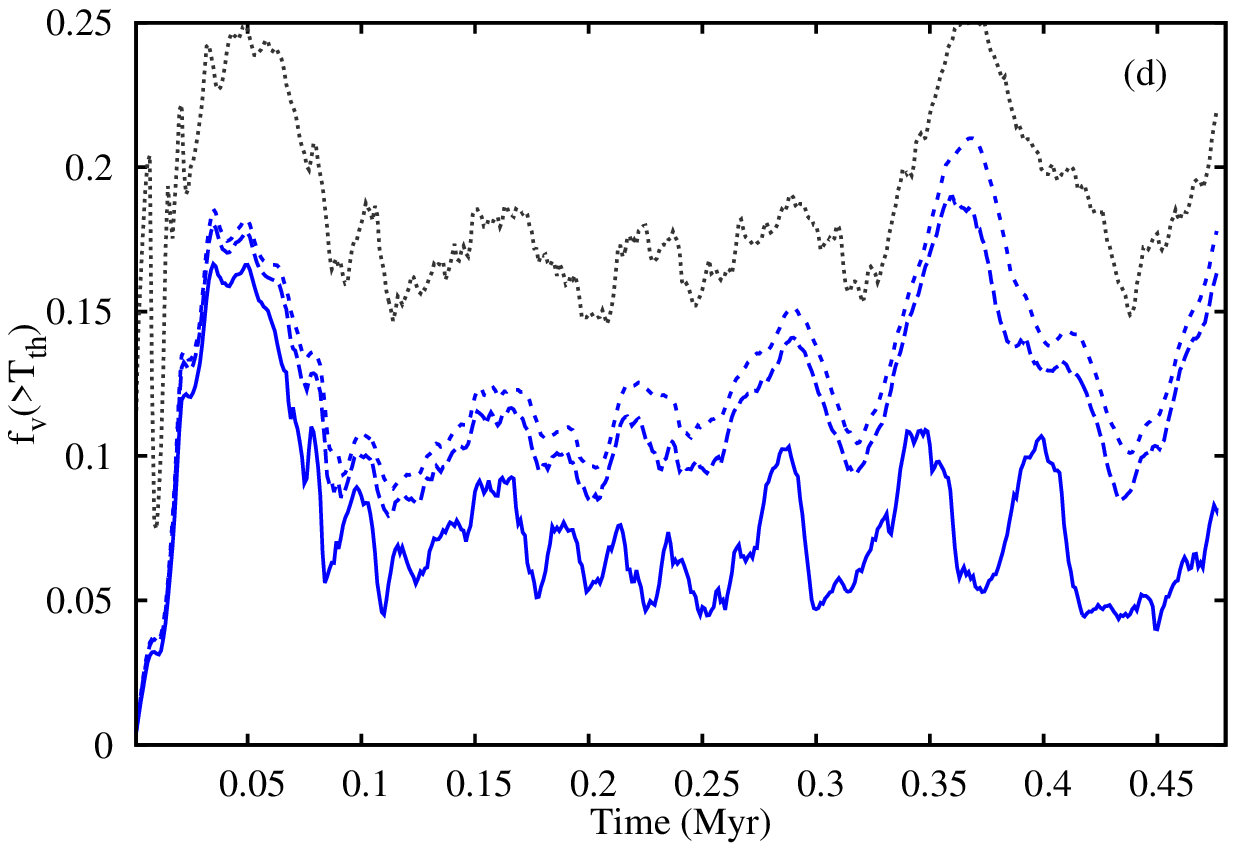}}
\caption{
  Filling factor in the H\,\textsc{ii} region of hot gas above the threshold temperatures indicated in the legend, as a function of time.
  Panel (a) is for simulation V16, (b) for V08, (c) for V06, and (d) for V04.
  The H\,\textsc{ii} region leaves the simulation domain downstream for simulations V16 and V08 at a time indicated with the vertical lines in panels (a) and (b).
  All photoionized gas in the H\,\textsc{ii} region that has not been shock-heated has $5000\,\mathrm{K} \leq T \leq 10\,000$ K.
  }
\label{fig:FillV}
\end{figure*}

\begin{figure}
\centering
\resizebox{0.9\hsize}{!}{\includegraphics{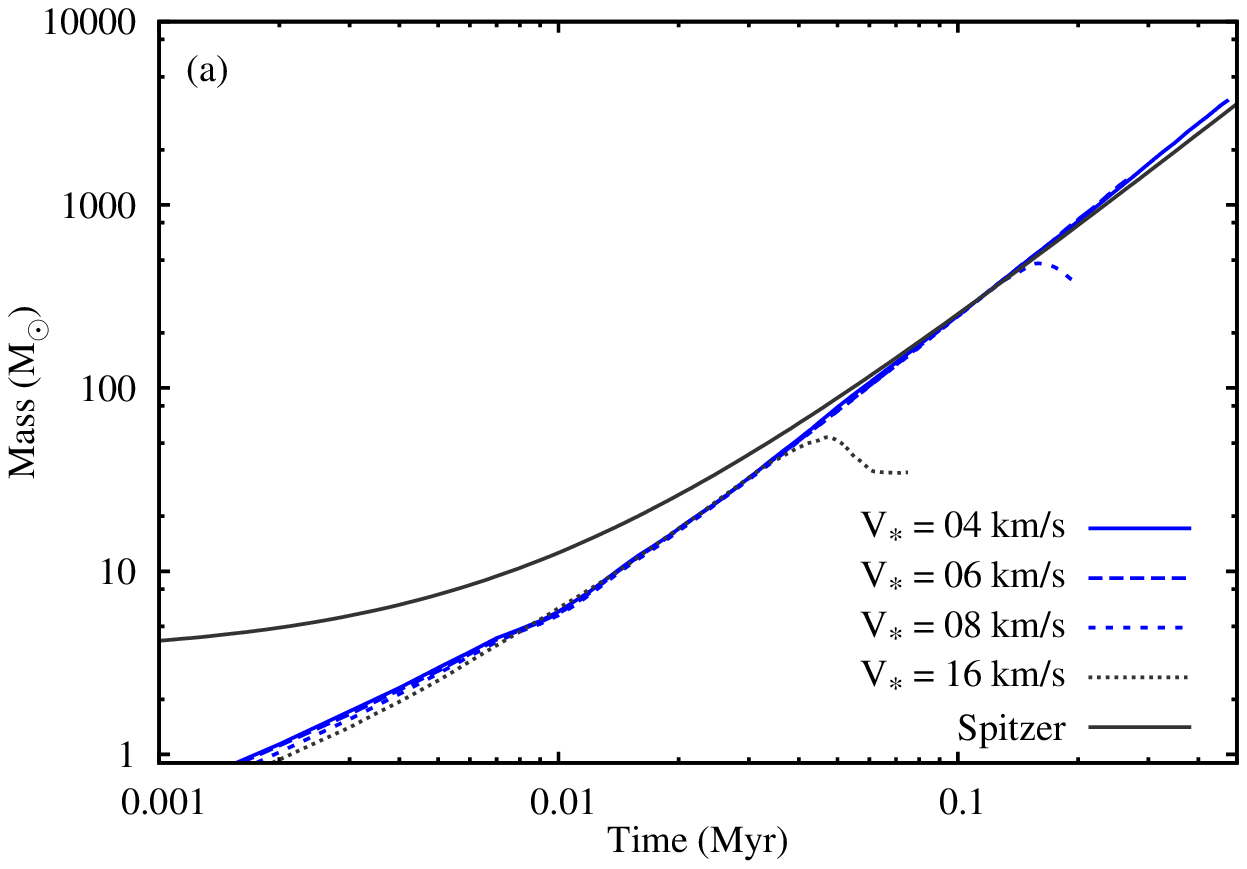}}
\resizebox{0.9\hsize}{!}{\includegraphics{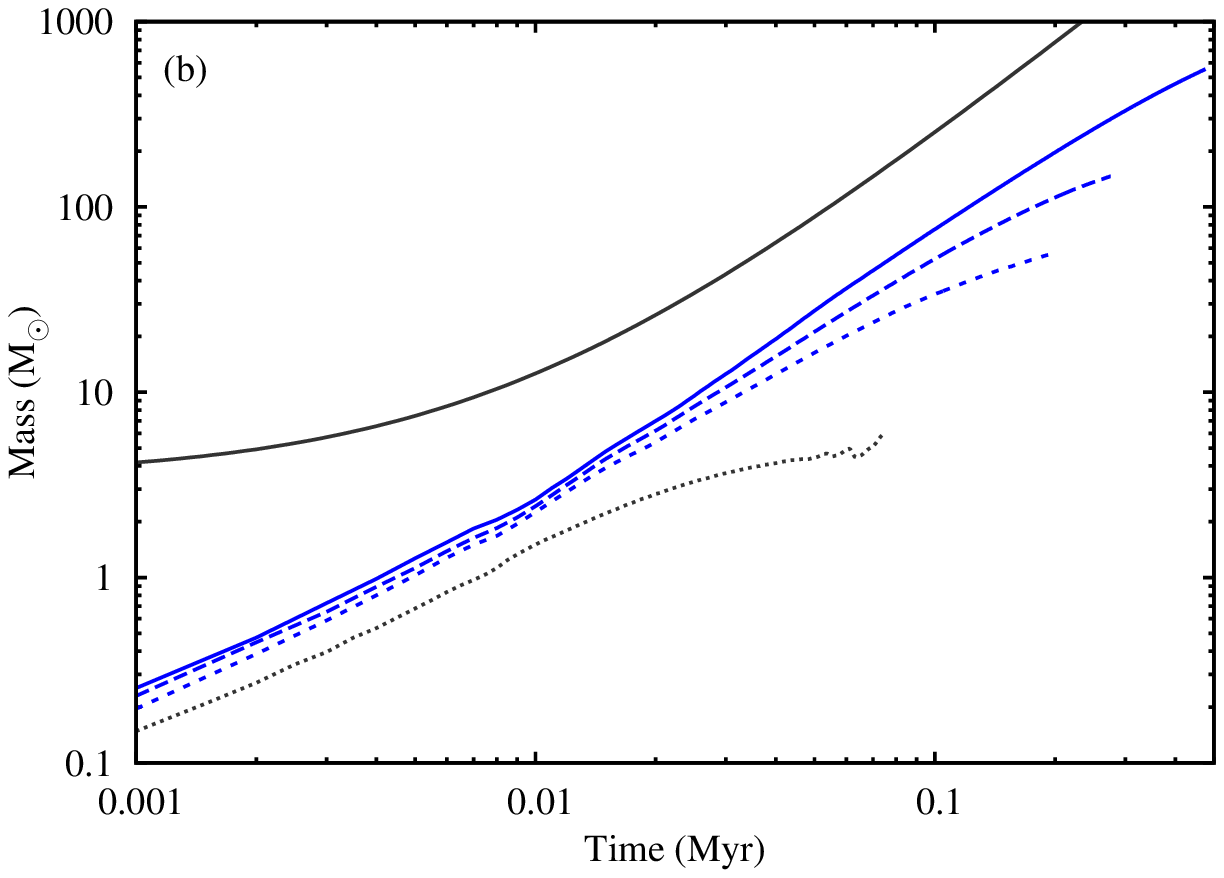}}
\caption{
  Mass of the H\,\textsc{ii} region shell for the simulations as a function of time.
  Panel (a) shows the mass in the full shell whereas panel (b) shows the mass in the part of the shell upstream from the star ($z>0$).
  In panel (a) the curves lie almost exactly on top of each other.
  }
\label{fig:shellmass}
\end{figure}

%%% ------------------------------------------------------------
\subsection{Properties of the wind and ISM bubbles}
%%% ------------------------------------------------------------

%%% ------------------------------------------------------------
\subsubsection{Filling factor of the hot bubble}
%%% ------------------------------------------------------------
Fig.~\ref{fig:FillV} shows the filling factor of hot gas above a threshold temperature, $f_\mathrm{v}(>T_\mathrm{th})$, defined as the ratio of volumes of gas with $T>T_\mathrm{th}$ and $T>5000$ K.
The latter is the H\,\textsc{ii} region volume because all photoionized gas in the simulation has $T>5000$ K.
For example, $f_\mathrm{v}(>10^6\,\mathrm{K})$ represents the fraction of the H\,\textsc{ii} region volume filled with gas with $T>10^6$ K.
For all simulations we find that the hot bubble with $T>10^7$ K is limited to 5-10 per cent of the H\,\textsc{ii} region volume, except for a transient phase in the early expansion of the bubble, before turbulent mixing sets in from waves reflected from the internal walls of the H\,\textsc{ii} region.
The differences between the four curves on each plot reflect the size of the mixing/cooling region where the thermal energy of the hot bubble is dissipated.
This region is dominated by numerical mixing in these simulations, so we cannot make strong statements about how realistic this is.
What we can say is that the wind bubble is always less than one quarter of the total volume of the H\,\textsc{ii} region, and nearly all of this volume is downstream from the star.
The animations of Figs.~\ref{fig:v16DT}-\ref{fig:v04DT} shows that the turbulent mixing is an intermittent process, driven by large KH rolls flowing downstream, and it is these rolls that drive the time variation in the hot bubble's filling factor.

%%% ------------------------------------------------------------
\subsubsection{Shell mass}
%%% ------------------------------------------------------------
The shell mass for the four simulations is plotted in Fig.~\ref{fig:shellmass} for the full shell in panel (a) and for just the upstream region ($z>0$) in panel (b).
The shell is defined as all grid zones with density more than 10 per cent above the background and with $x<0.1$, i.e.\ neutral gas that is overdense.
Panel (a) shows that all simulations sweep up the same mass in the H\,\textsc{ii} region shell.
This happens because all simulations have D-type ionization fronts in all directions, so the advective term just changes the location of the ionization front but not it's relative velocity through the ISM, at least to first order.
It is surprising, however, that the asymmetric internal structure of the H\,\textsc{ii} region does not affect the shell mass at late times.
The shell is advected out of the simulation domain at later times in V16 and V08, so their deviation is artificial.

We compare to the \citet{Spi78} solution, using $a_\mathrm{i}=12.5\,\mathrm{km}\,\mathrm{s}^{-1}$ (see discussion on H\,\textsc{ii} region radius below), assuming the ionized mass is negligible so that the shell mass is just the total mass swept up:
\begin{equation}
M_\mathrm{sh} = \frac{4}{3}\pi \rho_0 R_\mathrm{St}^3
\left( 1+ \frac{7 a_\mathrm{i} t}{4 R_\mathrm{St}} \right)^{12/7} \;.
\end{equation}
This overestimates the shell mass at early times because it assumes the H\,\textsc{ii} region mass is all swept into the shell, but then at later times the predicted shell mass agrees well with the analytic solution.
{\changed There may be some indication that the logarithmic slope of the predicted shell mass is shallower than is found in the simulation, but overall the agreement is very good.}
%This may be because the Spitzer solution calculates the I-front radius, whereas the outer edge of the shell is at a larger radius.
%Mass scales with radius cubed, so a moderately thick shell can easily account for a 50 per cent increase in mass.}

Panel (b) shows that in the upstream direction the shell mass does differ between the four simulations.
The difference consists of shell material that was upstream initially but has been advected downstream, and also gas that is photoionized from the inner wall of the shell.
The properties of the shell and ionization front do depend on $v_\star$, in particular because the ionization front is closer to the star for larger $v_\star$ and so the photoevaporation rate from the shell's inner wall is larger.

\begin{figure*}
\centering
\resizebox{0.45\hsize}{!}{\includegraphics{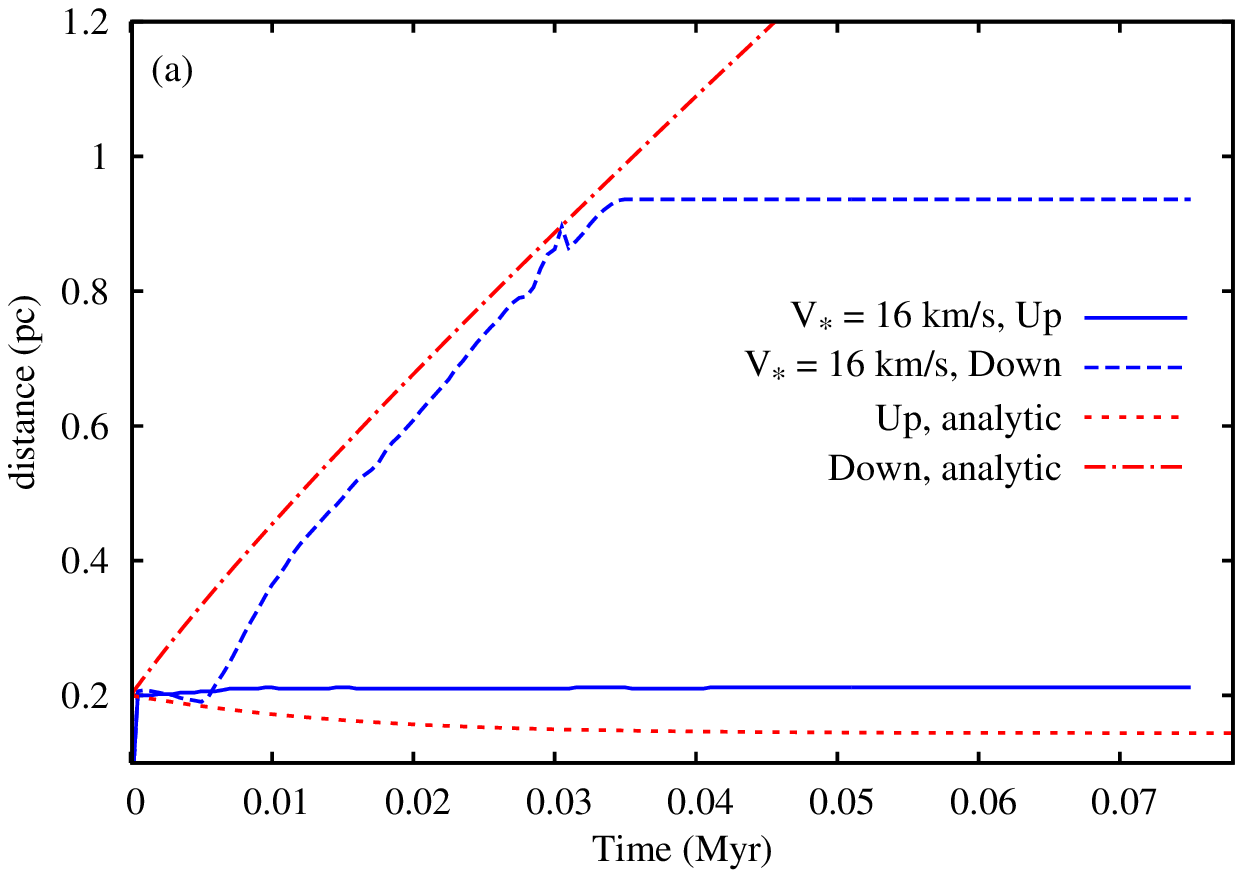}}
\resizebox{0.45\hsize}{!}{\includegraphics{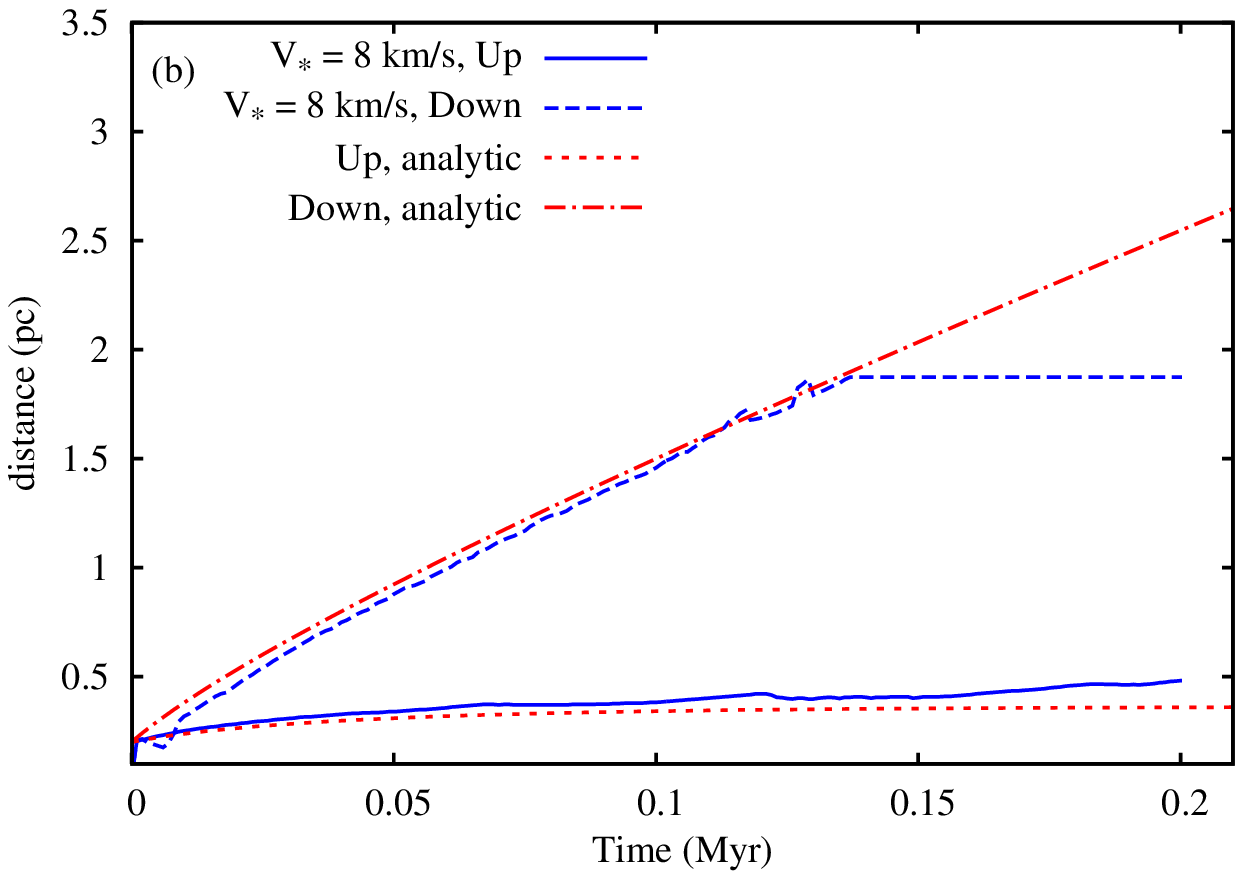}}
\resizebox{0.45\hsize}{!}{\includegraphics{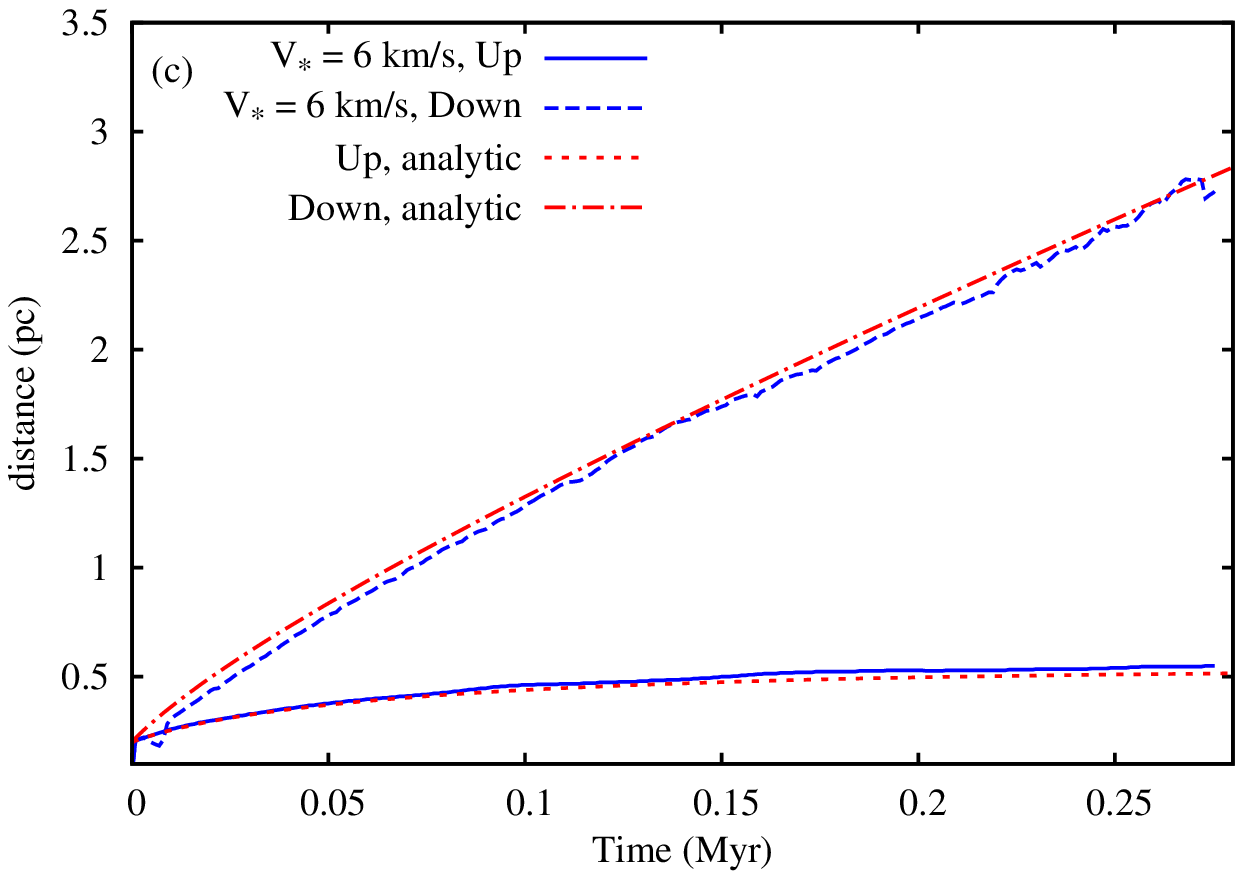}}
\resizebox{0.45\hsize}{!}{\includegraphics{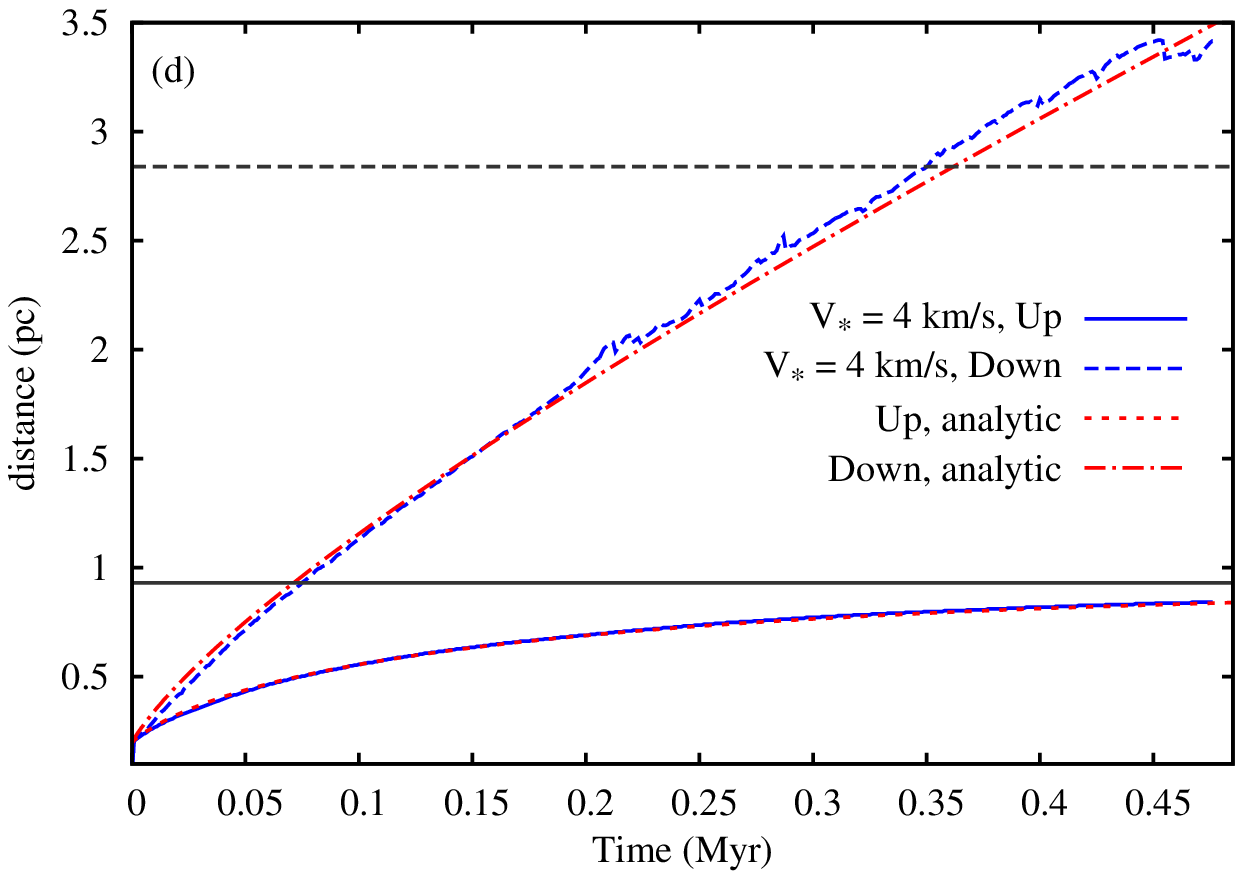}}
\caption{
  Upstream and downstream H\,\textsc{ii} region radius as a function of time.
  Panel (a) shows V16, (b) shows V08, (c) shows V06, and (d) shows V04.
  The blue lines show the numerical results, and the red lines show an analytic approximation obtained by adding an advection term to the usual H\,\textsc{ii} region expansion equation (see text for details).
  The horizontal lines at 0.93 pc and 2.84 pc in panel (d) show the observed upstream and downstream extent of the H\,\textsc{ii} region RCW\,120.
  }
\label{fig:updown}
\end{figure*}

%%% ------------------------------------------------------------
\subsubsection{H\,\textsc{ii} region radius}
%%% ------------------------------------------------------------
Fig.~\ref{fig:updown} plots the maximum upstream and downstream I-front position in the $z$ coordinate as a function of time for each simulation.
To compare to analytic expectations we take the \citet{Spi78} solution for the D-type expansion of H\,\textsc{ii} regions and add an advection term of velocity $v_\mathrm{ad}$, obtaining the differential equation for the I-front radius, $R_\mathrm{IF}$,
\begin{equation}
\dot{R}_\mathrm{IF}(t) = a_\mathrm{i}\left(\frac{R_\mathrm{st}}{R_\mathrm{IF}}\right)^{3/4}+v_\mathrm{ad} \,,
\end{equation}
where the dot denotes time derivative.
An analytic solution is no longer simple because separation of variables is not possible.
We have obtained a numerical solution with \emph{Mathematica} \citep{Wol91} using $v_\mathrm{ad}=-v_\star$ for the upstream I-front and $v_\mathrm{ad}=v_\star$ for the downstream I-front.
We could not get a good fit using $a_\mathrm{i}=9.87\,\mathrm{km}\,\mathrm{s}^{-1}$ (the sound speed in the interior of the H\,\textsc{ii} region) but the H\,\textsc{ii} region border is significantly warmer than the interior because of spectral hardening.
Increasing $a_\mathrm{i}$ to $12.5 \mathrm{km}\,\mathrm{s}^{-1}$ provided a much better fit so we used this for the analytic curves in Fig.~\ref{fig:updown}.

This simple extension to the Spitzer solution provides a reasonably good fit to the downstream I-front for all simulations.
The disagreement at early times is because the expanding wind bubble drives a compression wave that is initially quite dense and attenuates the ionizing photons.
All solutions then gradually relax to the advection velocity at late times; the evolution of V16 and V08 is cut off after 0.03 Myr and 0.14 Myr, respectively, when the downstream I-front exits the simulation domain.
For the upstream direction, V08, V06, and V04 are adequately fitted by the analytic approximation, with the agreement getting better for lower $v_\star$.
Simulation V16 shows disagreement however, because the Spitzer solution has a maximum expansion velocity of $a_\mathrm{i}$ at $R_\mathrm{IF}=R_\mathrm{st}$, whereas D-type I-fronts can actually propagate at up to $2a_\mathrm{i}$.
This shows the limitation of the Spitzer solution in describing the early phase expansion of H\,\textsc{ii} regions.
For simulation V04 we also plot the upstream and downstream dimensions of the H\,\textsc{ii} region RCW\,120.
At $t=0.35$ Myr we obtain a reasonable fit to the observations, although the upstream ionization front has not propagated quite far enough.

\begin{figure}
\centering
\resizebox{0.9\hsize}{!}{\includegraphics{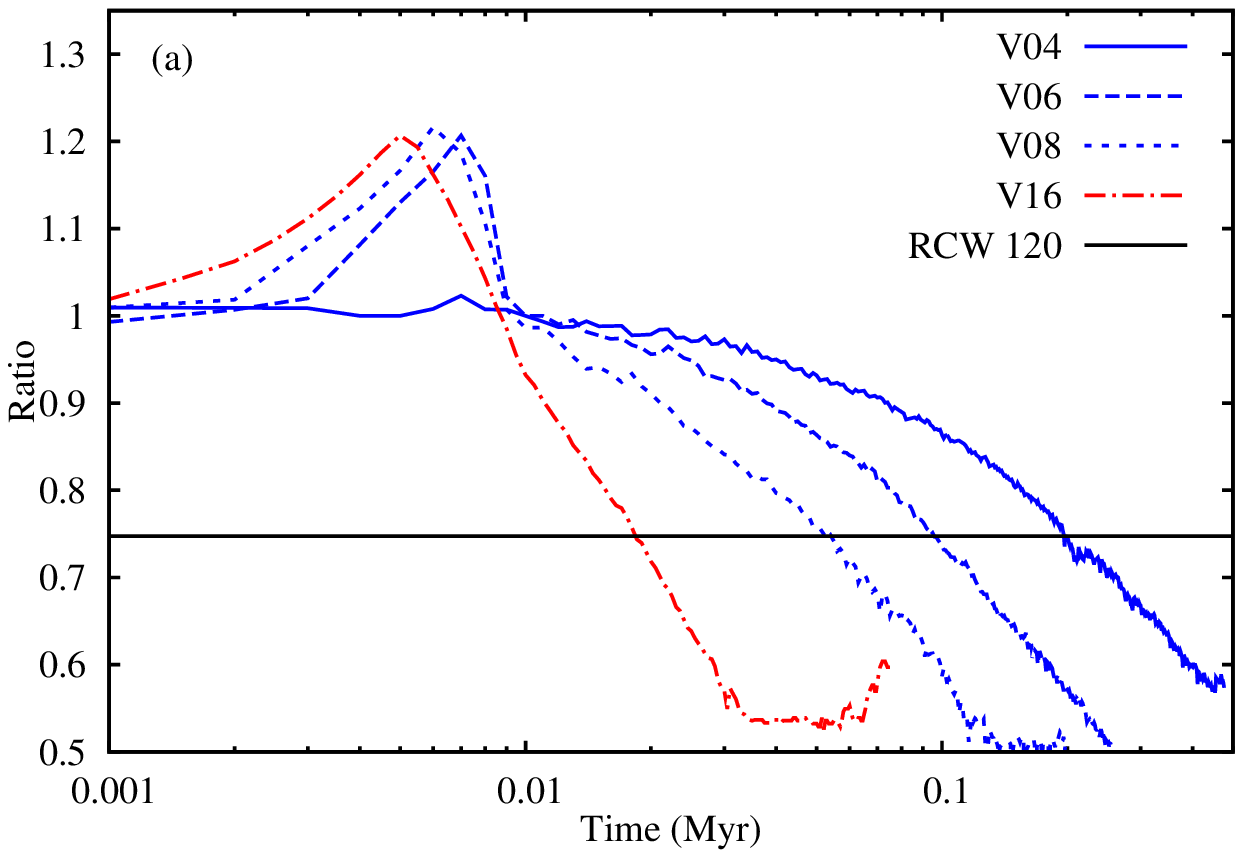}}
\resizebox{0.9\hsize}{!}{\includegraphics{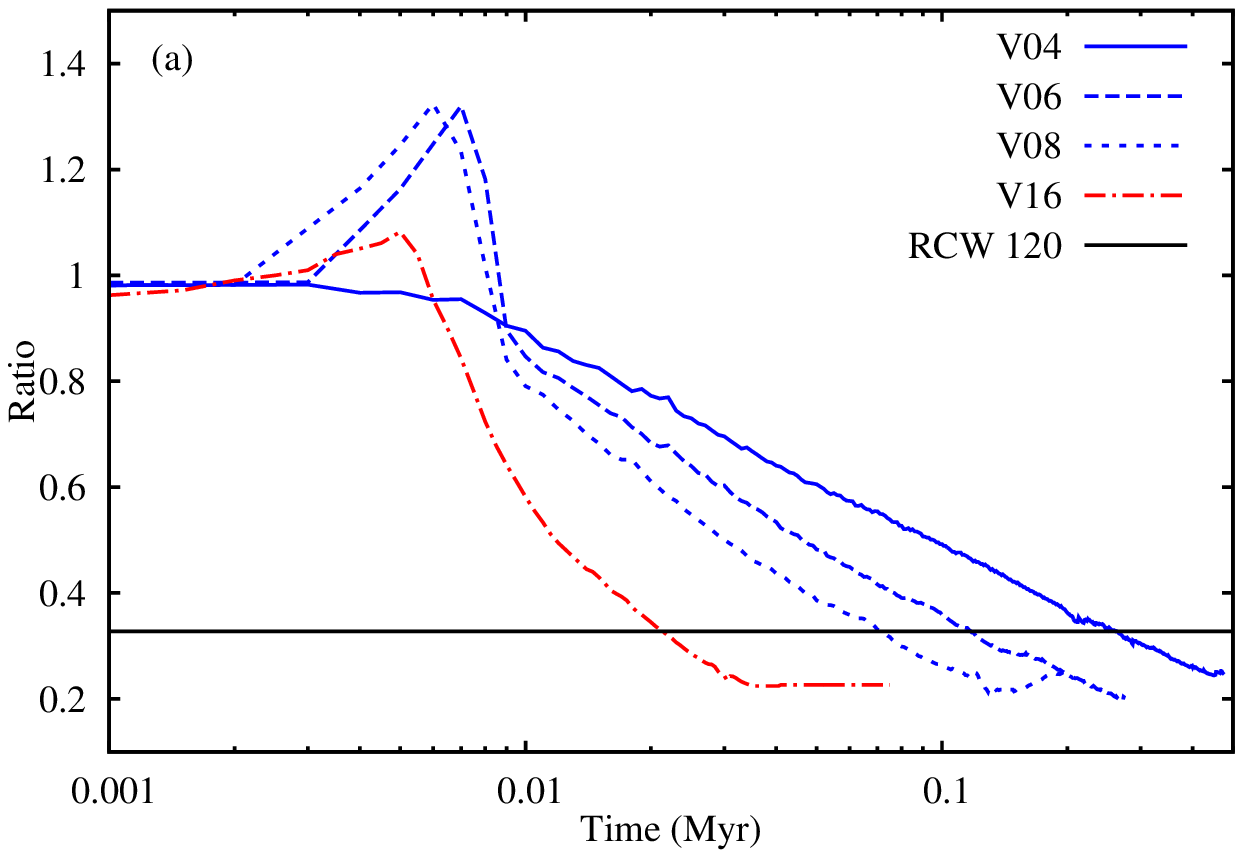}}
\caption{
  Size ratios of the H\,\textsc{ii} region shell for the simulations as a function of time.
  Panel (a) shows the breadth-to-length ratio and panel (b) shows the upstream-to-downstream ratio, both measured relative to the ionizing star's position.
  The horizontal lines show the observed ratios for the H\,\textsc{ii} region RCW\,120.
  }
\label{fig:ratios}
\end{figure}

The breadth-to-length ratio and upstream-to-downstream radius ratio are plotted in Fig.~\ref{fig:ratios} as a function of time for all four simulations.
None of the simulations has reached a steady state by the end of the simulation; when curves for V08 and V16 become horizontal it is because the H\,\textsc{ii} region has expanded beyond the simulation edges.
The initial increase in both ratios up to 10\,000 years is because the downstream H\,\textsc{ii} region shrinks (because of the aforementioned compression wave driven by the expanding wind bubble).
This transient feature disappears once the wind bubble's expansion becomes subsonic in the downstream direction.
Thereafter both ratios decrease over time, with the rate being proportional to $v_\star$ because this determines the distortion of the bubble from sphericity.
Both ratios cross the observed values for RCW\,120 at $t\approx0.25$ Myr for simulation V04, at an earlier time than that for which the absolute sizes match the observations (0.35 Myr).

%%% ------------------------------------------------------------
\subsection{Flow of gas through the H\,\textsc{ii} region}
%%% ------------------------------------------------------------

\begin{figure}[h]
\centering
\resizebox{1.0\hsize}{!}{\includegraphics{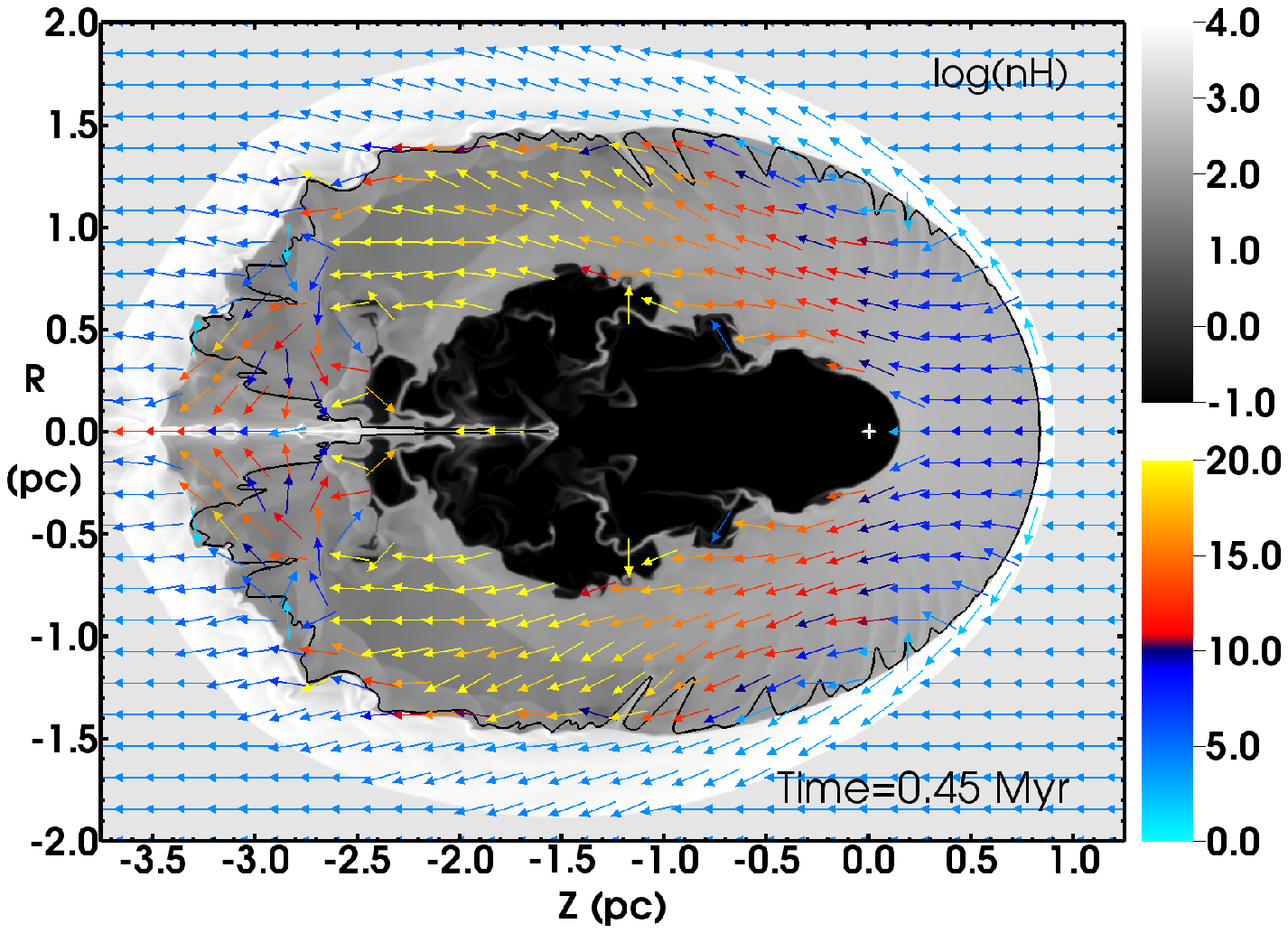}}
\caption{
  Log of $n_{\mathrm{H}}{}$ (greyscale) and gas velocity relative to the star (coloured arrows) for simulation V04, with units $\log(n_{\mathrm{H}}{}/\mathrm{cm}^{-3})$ and $\mathrm{km}\,\mathrm{s}^{-1}$, respectively.
  The radial direction is vertical, and the axis of symmetry is $R=0$, the $z$-axis.
  The solid black contour shows $x=0.5$.
  The star is at the origin, denoted by a white cross.
  Regions with gas velocity $>30\,\mathrm{km}\,\mathrm{s}^{-1}$ are excluded from the velocity plot for clarity, and velocities from 20 to 30\,km\,s$^{-1}$ all have the same colour arrows (yellow).
  An animation of the simulation's evolution is available online.
  }
\label{fig:v04DV}
\end{figure}

Fig.~\ref{fig:v04DV} shows the gas flow through simulation V04 after 0.45 Myr of evolution.
The H\,\textsc{ii} region shell deflects much of the upstream ISM away from the star, but interestingly the interior of the H\,\textsc{ii} region shows quite strong gas acceleration from upstream to downstream.
This is very similar to the Champagne flow model \citep{Ten79} because, as discussed above, the moving star H\,\textsc{ii} region also has a pressure gradient.
At $z=0$ the gas is already flowing past the star at $v\approx10\,\mathrm{km}\,\mathrm{s}^{-1}$ (meaning it has accelerated by $6\,\mathrm{km}\,\mathrm{s}^{-1}$), and far downstream it is moving at $v>20\,\mathrm{km}\,\mathrm{s}^{-1}$ before it hits the downstream wall of the H\,\textsc{ii} region.
In principle this acceleration of the gas could cause the formation of a wind bow shock, even for stars moving subsonically with respect to the undisturbed ISM.

%%% ------------------------------------------------------------
\subsection{X-ray emission}
%%% ------------------------------------------------------------

\begin{figure}
\centering
\resizebox{0.8\hsize}{!}{\includegraphics{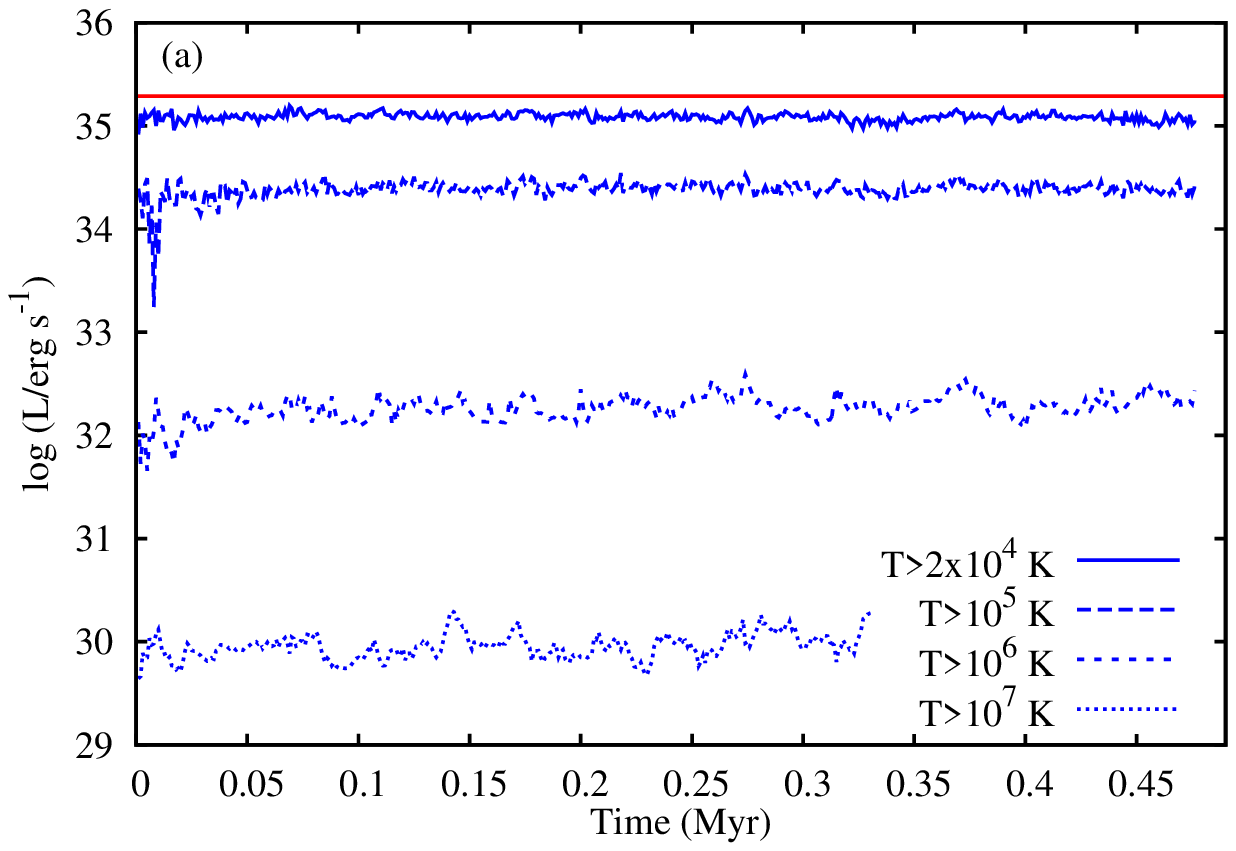}}
\resizebox{0.8\hsize}{!}{\includegraphics{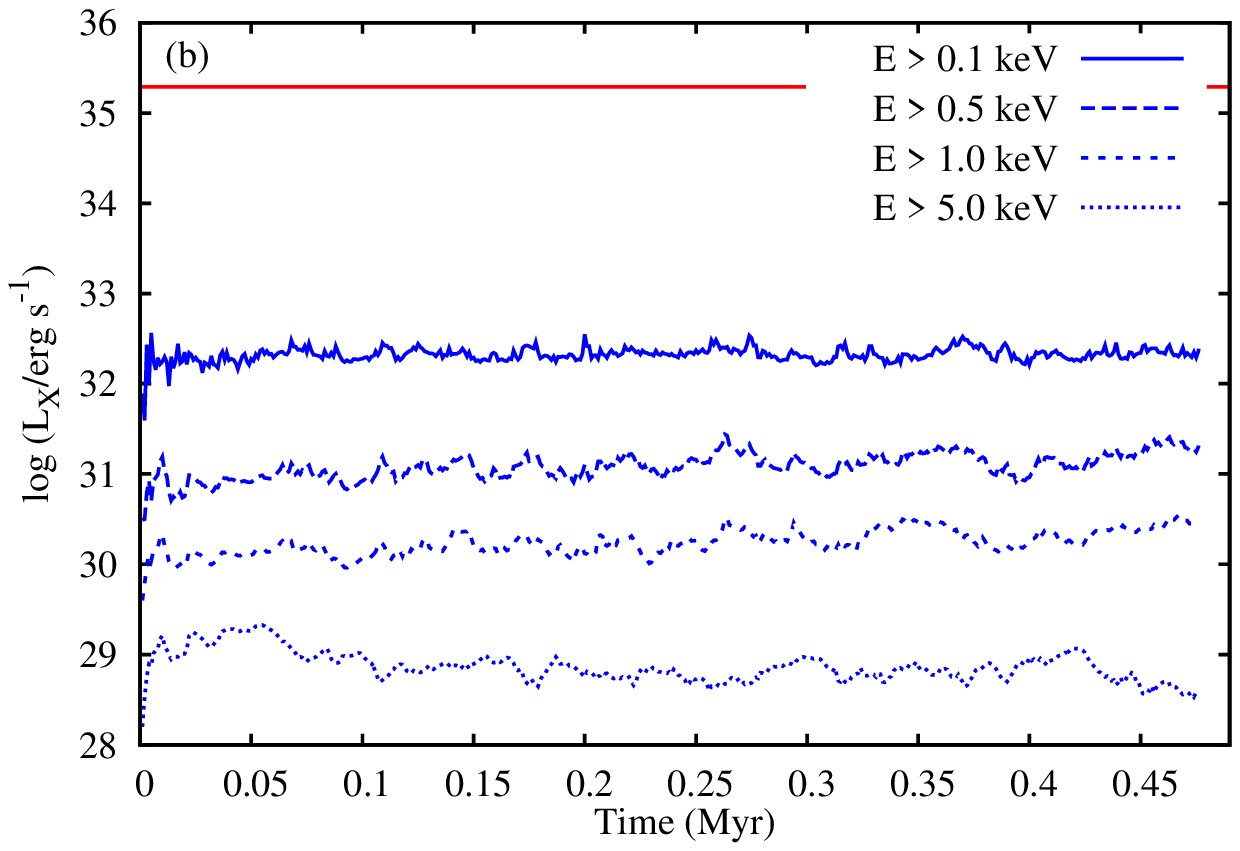}}
\caption{
  Total and X-ray cooling from the stellar wind bubble of simulation V04 as a function of time.
  Panel (a) shows the total cooling luminosity of all gas with $T>2\times10^4$, $10^5$, $10^6$, and $10^7$ K (for the lower temperatures most of this is not in at X-ray wavelengths).
  Panel (b) shows the total (unattenuated) X-ray luminosity above energies $E>0.1$, 0.5, 1.0, and 5.0 keV, calculated assuming collisional ionization equilbrium and solar abundances (see text for details).
  In both panels the red horizontal line shows the mechanical luminosity of the stellar wind, $0.5\dot{M}v_\mathrm{w}^2$.
}
\label{fig:xrayV04}
\end{figure}

The stellar wind mechanical luminosity (i.e.\ energy input rate to the wind bubble) is $L_\mathrm{w}=0.5\dot{M}v_\mathrm{w}^2=1.95\times10^{35}\,$erg\,s$^{-1}$.
Some fraction of this goes into driving the expansion of the bubble, and the rest is radiated away.
If there were no mixing at the bubble's edge, then all of this radiation would be in X-rays because the bubble is so hot.
In Fig.~\ref{fig:xrayV04} we plot the radiative luminosity of the hot gas in simulation V04 as a function of time through the simulation; results from the other simulations are very similar.
Panel (a) shows the total cooling luminosity from gas above threshold temperatures $T=2\times10^4$, $10^5$, $10^6$, and $10^7$ K, using the cooling curve described in Sect.~\ref{ssec:cooling}.
All of the gas with $T>2\times10^4$ K is either stellar wind material or has mixed with stellar wind material and been heated through this mixing.
This explains why the total luminosity of this gas {\changed almost matches} the mechanical input luminosity of the stellar wind, i.e., {\changed most} of the stellar wind's kinetic energy is eventually radiated away.
We see, however, that only about 0.1 per cent of this ($L\sim(2-3)\times10^{32}$ erg\,s$^{-1}$) is radiated away by gas with $T>10^6$ K.
The rest of the energy is radiated away by colder gas in the mixing region, which will not be emitted as X-rays.

For $T\gtrsim10^7$ K most of the X-ray emission is thermal Bremsstrahlung, but for lower temperatures line emission dominates \citep[e.g.,][]{RogPit14}.
We used the \textsc{xspec v12} package \citep{Arn96} to generate a table of the X-ray emissivity as a function of temperature for photon energy thresholds, $E>0.1$, 0.5, 1.0, and 5.0 keV, using the \textsc{apec} model with solar metallicity \citep{AspGreSauEA09}.
Interpolation of these tables allowed us to estimate the X-ray luminosity, $L_\mathrm{X}$, of each snapshot of the simulations, and this is shown in panel (b) of Fig.~\ref{fig:xrayV04}, again for simulation V04.
Above 0.1 keV, $L_\mathrm{X}$ is very similar to the total cooling luminosity for gas with $T>10^6$ K, because 0.1 keV corresponds to $1.16\times10^6$ K.
For higher energy thresholds the luminosity decreases substantially.
Again, {\changed at most 0.1 per cent} of the wind mechanical luminosity is radiated in X-rays.
This figure shows very clearly how turbulent mixing (and numerical heat conduction) efficiently remove most of the wind bubble's energy, and how the lower-temperature mixed gas then radiates away the energy in optical and UV lines.
This does not significantly change the energy budget of the H\,\textsc{ii} region because the wind mechanical luminosity is only about 1 per cent of the photoheating rate from photoionization.

The predicted X-ray luminosities are significantly smaller than the luminosities detected for Wolf-Rayet bubbles, $L_\mathrm{X}\gtrsim10^{33}$ erg\,s$^{-1}$ \citep[see e.g.,][]{ToaGue13, DwaRos13, ToaGueGruEA14}, because the O star wind is much weaker.
The discrepency is actually even larger because these are observed luminosities, significantly attenuated by line-of-sight absorption.
Young H\,\textsc{ii} regions such as RCW\,120 generally have significant extinction along the line-of-sight so the soft X-rays ($0.1-0.5$ keV), where the bubble emits most strongly, would be significantly absorbed.

Our predicted unattenuated luminosity is consistent with the wind bubbles simulated by \citet{RogPit14}; they predict a luminosity $L_\mathrm{X}\gtrsim10^{33}$ erg\,s$^{-1}$ during the first 0.5 Myr of their simulation, but they also consider more massive stars with a combined wind mechanical luminosity about 7 times larger than in our calculations.
Similarly, the winds from main sequence stars in three-dimensional superbubble simulations by \citet{KraDieBohEA14} are significantly stronger than in our models, and the X-ray emission they predict is consequently larger ($L_\mathrm{X}\sim3\times10^{33}$ erg\,s$^{-1}$).
They found that only a few times $10^{-4}$ of the input mechanical energy is radiated in X-rays, slightly below but comparable to the fraction that we find.
Compared to the simulations of \citet{ToaArt11}, we find a larger X-ray luminosity for a lower-mass star (they find $L_\mathrm{X}\approx 5\times 10^{31}$ erg\,s$^{-1}$ for the main sequence bubble around a 40 \msun{} star), but this may be explained by the much higher density and pressure medium that we consider.

We find that the soft X-ray ($<1$ keV) emission is strongly limb-brightened because it is emitted almost entirely from the mixing region between the wind bubble and the H\,\textsc{ii} region.
Only the  hard ($>5$ keV) emission appears as a filled bubble, although even it is somewhat limb-brightened and is also significantly brighter upstream from the star than downstream.
This morphology is very similar to that measured by \citet{ChuGueGruEA03} for the Wolf-Rayet nebula S308, where the X-ray emission is limb-brightened and dominated by gas with $T\sim10^6$ K, with a possible small contribution from hotter gas.

%%% ------------------------------------------------------------
%%% ------------------------------------------------------------
\section{Discussion}
\label{sec:discussion}
%%% ------------------------------------------------------------
%%% ------------------------------------------------------------

%%% ------------------------------------------------------------
\subsection{Mixing at the contact discontinuity}
%%% ------------------------------------------------------------
\citet{RosLopKruEA14} compared the energy input from stellar winds to the ISM around massive star clusters, finding very little of the energy is radiated away in X-rays and also that only a fraction goes into work done driving expansion of the bubble.
They speculate that much of the energy is dissipated by turbulent mixing of the hot shocked wind with the cooler ISM, and resultant cooling through line emission (or alternatively by heat transport through thermal conduction from the wind to the ISM).
Our results support this picture, albeit on the much smaller scale of a wind bubble around a single star.
The mixing is driven largely by Kelvin-Helmholz instability at the contact discontinuity, from shear flows that arise because the sound speeds of the two phases differ by a factor of about 100.
One should bear in mind, however, that vortices have very different properties in 2D compared with 3D.
Also, we do not include magnetic fields or thermal conduction in this work, so the structure of the contact discontinuity is determined by numerical diffusion and not by physical processes.
Thermal conduction can, in the absence of magnetic fields, strongly modify the contact discontinuity and shocked wind region in stellar wind bubbles \citep{ComKap98, MeyMacLanEA14}.
We therefore do not draw strong conclusions about turbulent mixing from the results of the simulations presented here.
On the other hand, the X-ray morphology that we predict is rather similar to that observed for the more luminous WR bubble S308 \citep{ChuGueGruEA03}, so it is worthwhile to explore the turbulent mixing process with more detailed future simulations.

%%% ------------------------------------------------------------
\subsection{Comparison to RCW\,120}
\label{ssec:RCW120}
%%% ------------------------------------------------------------

\begin{table}
  \centering
  \caption{
    Comparison of the H\,\textsc{ii} region RCW\,120 with simulation V04 at three different times (in Myr).
    The dimensions of RCW\,120 are measured from the \emph{Spitzer} data assuming a distance of 1.35 kpc \citep{MarPomDehEA10}, and the shell mass estimate is from \citet{DehZavSchEA09}.
  }
  \begin{tabular}{ l l l l l}
    %\hline
    Quantity  &  $t=0.4$ & =0.35 & =0.25  & RCW\,120 \\
    \hline
    Length (pc) & 3.97 & 3.64 & 2.97 & 3.76 \\
    Breadth (pc) & 2.87 & 2.72 & 2.38 & 3.41 \\
    Upstream offset (pc) & 0.82 & 0.80 & 0.74 & 0.93 \\
    Downstream offset (pc) & 3.15 & 2.84 & 2.23 & 2.84 \\
    Perpendicular offset (pc) & 1.19 & 1.15 & 1.04 & 1.40 \\
    Shell mass ($\msun$) & 5600 & 4400 & 2400 & 1200-2100 \\
  \end{tabular}
  \label{tab:RCW120}
\end{table}

\citet{ArtHenMelEA11} compared their simulations of H\,\textsc{ii} region expansion in a turbulent ISM to RCW\,120, estimating the age of the H\,\textsc{ii} region to be $\approx0.2$ Myr based on its size and the mass of its shell.
In contrast, Table~\ref{tab:RCW120} shows that our simulations have shell masses that are too large for RCW\,120 at times $0.25-0.4$ Myr, even though the H\,\textsc{ii} region is still too small.
The ratio of shell mass to shell radius depends only on the mean gas density, suggesting that the mean density in our simulations is somewhat too large for direct comparison.
Indeed \citet{ArtHenMelEA11} estimated the mean number density to be $n_0\approx1000\,\mathrm{cm}^{-3}$, whereas we are using a density three times larger, based on the upper limit to the ISM density from \citet{ZavPomDehEA07}.
This also explains why our simulations suggest a larger age for RCW\,120 than what \citet{ArtHenMelEA11} obtained, because H\,\textsc{ii} regions expand more slowly in a higher density medium.
In any case, we can conclude that the H\,\textsc{ii} region should be $\lesssim 0.4$ Myr old.

\citet{OchVerCoxEA14} proposed a model in which a strong Champagne flow has been established in the H\,\textsc{ii} region of RCW\,120 over 2.5\,Myr.
While an age of 2.5\,Myr could be consistent with the age of the star, which is only constrained to be $<3$\,Myr old \citep{MarPomDehEA10}, it is not consistent with the mass of the swept-up shell.
\citet{OchVerCoxEA14} find that the Champagne flow would be much weaker at $t=0.5$\,Myr, so it is not clear if their model will still work at this earlier time.

Three different models have now been proposed for RCW\,120:
\citet{ArtHenMelEA11} consider H\,\textsc{ii} region expansion around a static star in a turbulent magnetised ISM, \citet{OchVerCoxEA14} propose a Champagne flow explanation, and we have considered stellar motion through the ISM as a possible source of the H\,\textsc{ii} region and wind bubble asymmetry.
It is of course likely that elements of all three models are present in reality, because molecular clouds are turbulent, have large-scale density gradients, and give birth to slowly moving stars.
Quantitative comparison of the ionized gas and shell kinematics with simulations may be required to determine whether one source of asymmetry is dominant over others.

Our simulations show quite clearly that CD\,$-$38$\deg$11636 cannot have a large space velocity, and we can limit it to $v_\star\lesssim4\,\mathrm{km}\,\mathrm{s}^{-1}$.
Such a small velocity is most likely to arise for a star that formed in situ from turbulent initial conditions in the molecular cloud.
There should be a Champagne flow occuring at some level (molecular clouds are not uniform on parsec length scales) so it is probable that $0\,\mathrm{km}\,\mathrm{s}^{-1} \leq v_\star \lesssim 4\,\mathrm{km}\,\mathrm{s}^{-1}$.
This $30\,\msun$ star is apparently the only ionizing source in RCW\,120 \citep{MarPomDehEA10}, and most of the identified young stars are found $\gtrsim1$ pc away in the H\,\textsc{ii} region shell \citep{ZavPomDehEA07, DehZavSchEA09}.
It is not clear if this region has intermediate-mass stars, but it may be very interesting to characterise the stellar mass function in this region.
This would show whether the presence of such a massive star in apparent isolation from other massive stars conforms to expectations from theoretical models \citep[e.g.,][]{WeiKroPfl13} and simulations \citep[e.g.,][]{PetBanKleEA10, DalBon11}.

It would be very interesting to make IR maps from our simulations to compare to the \emph{Spitzer} and \emph{Herschel} observations of RCW\,120 \citep{ZavPomDehEA07, DehZavSchEA09, AndZavRogEA10}.
Unfortunately this is a rather complicated process.
\citet{PavKirWie13} showed that the 8\,$\mu$m emission from PAH molecules cannot be modelled assuming there is no processing within the H\,\textsc{ii} region.
To get the ring-like emission (see Fig.~\ref{fig:rcw120}) they needed to destroy the PAH particles in the H\,\textsc{ii} region interior, otherwise the 8\,$\mu$m and 24\,$\mu$m emission have similar spatial distributions.
If the 24\,$\mu$m emission comes from very small grains, then a central cavity evacuated by the stellar wind could explain the observations.
\citet{OchVerCoxEA14} argue that a cavity created by radiation pressure on dust grains, excluding the grains from the vicinity of the ionizing star, could also produce the 24\,$\mu$m emission.
In light of these complications, it is beyond the scope of this paper to predict the IR emission from our simulations, but we intend to pursue this in future work.

%%% ------------------------------------------------------------
\subsection{Do any stellar wind bubbles fill their H\,\textsc{ii} regions?}
%%% ------------------------------------------------------------
The bubble N49 was modelled by \citet{EveChu10}; they found that the interior of the H\,\textsc{ii} region cannot be a pure wind bubble because of its dust content.
They conclude it must at least contain a mixture of ISM and wind material, following the model of \citet{McKVanLaz84}.
The double-shell structure of N49 found by \citet{WatPovChuEA08} is, however, suggestive of an inner wind bubble separated from a larger H\,\textsc{ii} region by a boundary layer of mixed wind and ISM.
In this case, as for RCW\,120, it is difficult to unambiguously state what the filling fraction of the wind bubble is, and how well-mixed the wind material is with the ISM. 
There is, as far as we are aware, no clear observational evidence for any single O star that its wind bubble completely fills its H\,\textsc{ii} region.
Our results support the picture obtained in previous work \citep{FreHenYor03,FreHenYor06,ToaArt11} that winds from mid-to-late O stars are not strong enough to drive such large wind bubbles.
Recent results from simulations of star cluster formation \citep{DalNgoErcEA14} show that also in this case the wind cavity remains smaller than, and distinct from, the H\,\textsc{ii} region.

Very massive stars may be rather different, however.
A spectacular example is VFTS\,682 \citep{BesVinGraEA11}, with stellar mass $M_\star\approx150\,\msun$ and located a projected distance of 29 pc from the star cluster R\,136 in the Large Magellanic Cloud.
In the absence of proper motion measurements for this star \citet{BesVinGraEA11} were unable to decide if the star is an exile \citep[probably ejected from R\,136;][]{BanKroOh12, GvaWeiKroEA12} or could have formed in situ.
Its mass-loss rate is $\dot{M}\approx2\times10^{-5}\,\msunperyr$ and its ionizing photon luminosity is $Q_0\approx2\times10^{50}\,$s$^{-1}$ \citep{BesGraVinEA14}, both about 100 times larger than what we have simulated.
Preliminary calculations suggest that the shell around the wind bubble driven by this star would trap its H\,\textsc{ii} region within about $10^5$ years, if located in a similarly dense medium to the simulations in this work.
Similarly, \citet{ArtHoa06} found that the wind bubble shell around a star with $\dot{M}\approx10^{-6}\,\msunperyr$ traps its H\,\textsc{ii} region, and \citet{Ver14} find a similar result for an ultra-compact H\,\textsc{ii} region around a star with $\dot{M}=7\times10^{-6}\,\msunperyr$.

\citet{WeaMcCCasEA77} estimate (based on the column density of gas swept up by the wind bubble) that the H\,\textsc{ii} region will be trapped if $L_{36}^3 n_0 Q_{48}^{-2} \gtrsim 0.005$, where $L_{36}$ is the wind mechanical luminosity ($0.5\dot{M}v_\mathrm{w}^2$) in units of $10^{36}$\,erg\,s$^{-1}$ and $Q_{48}=Q_0/10^{48}\,$s$^{-1}$.
According to this criterion the wind bubble we simulate should trap the H\,\textsc{ii} region ($L_{36}^3 n_0 Q_{48}^{-2}\approx1.0$), but this is somewhat misleading because we only simulate the early expansion phase and not the steady state solution.
Perhaps eventually after a few million years the wind bubble would indeed fill the H\,\textsc{ii} region completely.
This criterion shows that the important ratio is $\dot{M}^3/Q_0^2$ which, for example, is 10 times larger for VFTS\,682 than for CD\,$-$38$\deg$11636.
Similarly, the strong wind simulations of \citet{ArtHoa06} and \citet{Ver14} have a much larger $\dot{M}^3/Q_0^2$ than what we simulate.

%%% ------------------------------------------------------------
%%% ------------------------------------------------------------
\section{Conclusions}
\label{sec:conclusions}
%%% ------------------------------------------------------------
%%% ------------------------------------------------------------
We have investigated the simultaneous expansion of a stellar wind bubble and H\,\textsc{ii} region around a single O star moving slowly through a dense, uniform ISM.
Our two-dimensional radiation-hydrodynamics simulations show that the stellar wind bubble is asymmetric from the star's birth, whereas the H\,\textsc{ii} region takes much longer to respond to stellar motion because of its much lower temperature (and sound speed).
Stellar wind bubbles fill about 10-20 per cent of the H\,\textsc{ii} region volume, and their shape is compatible with the interpretation that mid-IR arcs of dust emission seen in some H\,\textsc{ii} regions represent the upstream boundary between the wind bubble and the photoionized ISM.

The expansion rate of the H\,\textsc{ii} region for $v_\star\leq a_\mathrm{i}$ can be understood by a simple extension to the \citet{Spi78} solution including an advection term, although this seems to fail for $a_\mathrm{i}<v_\star<2a_\mathrm{i}$, where the D-type I-front advances with velocity $>a_\mathrm{i}$ with respect to the neutral unshocked gas.
The shell mass around the H\,\textsc{ii} region appears insensitive to stellar motion, as long as $v_\star<2a_\mathrm{i}$ (i.e.\ as long as a shell can form).
The internal dynamics of the photoionized gas in the H\,\textsc{ii} region develops similarly to Champagne flows in static star H\,\textsc{ii} regions, in that the pressure asymmetry drives acceleration from upstream to downstream.
This is not strongly affected by the presence of the stellar wind because the wind bubble occupies such a small fraction of the H\,\textsc{ii} region volume.

The simulation V16 has a stellar wind bow shock because it is moving with $v_\star>a_\mathrm{i}$.
The overdense bow shock absorbs some of the stellar ionizing photons, changing the shape of the H\,\textsc{ii} region somewhat.
There is a kink in the H\,\textsc{ii} region in the direction perpendicular to stellar motion, because here the path length through the bow shock is longest and part of the bow shock can trap the I-front.
In the direction of motion the H\,\textsc{ii} region is not significantly affected by the bow shock, and the two structures remain distinct with the H\,\textsc{ii} region radius about three times the bow shock radius.

The wind bubble has soft and faint X-ray emission, which should be limb-brightened because it arises mainly from the turbulent layer at the bubble's edge where wind and ISM material mix.
Very little of the kinetic energy input from the stellar wind is radiated as X-rays  ($<1$ per cent).
Most of the energy is radiated by cooler gas in the mixing layer, which will be observed as optical and ultraviolet spectral lines from metals.
This supports recent work \citep{RosLopKruEA14} where it was argued that thermal conduction or turbulent mixing of wind and ISM gas is the dominant energy loss mechanism of stellar wind bubbles.

Comparison of our simulations with the H\,\textsc{ii} region RCW\,120 shows that its dynamical age is $\lesssim0.4$ Myr and that stellar motion of $\lesssim4\,\mathrm{km}\,\mathrm{s}^{-1}$  is allowed (although the star may also be static), implying that the driving star CD\,$-$38$\deg$11636 probably formed in situ and is unlikely to be a runaway star.
In future work we will use radiative transfer postprocessing of these simulations to make synthetic dust and line emission maps to compare with the wealth of infrared data on RCW\,120 and other similar young interstellar bubbles.

%%% ------------------------------------------------------------
%%% ------------------------------------------------------------
\begin{acknowledgements}
JM acknowledges funding from from the Alexander von Humboldt Foundation for this work.
This project was supported by the the Deutsche Forschungsgemeinschaft priority program 1573, Physics of the Interstellar Medium.
VVG acknowledges the Russian Science Foundation grant 14-12-01096.
SM gratefully acknowledges the receipt of research funding from the National Research Foundation (NRF) of South Africa.
The authors gratefully acknowledge the computing time granted by the John von Neumann Institute for Computing (NIC) and provided on the supercomputer JUROPA at J\"ulich Supercomputing Centre (JSC).
The authors gratefully acknowledge the Gauss Centre for Supercomputing e.V. (\url{www.gauss-centre.eu}) for funding this project by providing computing time on the GCS Supercomputer SuperMUC at Leibniz Supercomputing Centre (LRZ, \url{www.lrz.de}) (project pr85jo).
JM is grateful to Vijaysarathy Bharadwaj for advice on the usage of \textsc{xspec}.
The authors thank the referee for useful comments that improved the paper.
\end{acknowledgements}
%%% ------------------------------------------------------------
%%% ------------------------------------------------------------

\bibliographystyle{aa}
\bibliography{../../../../../documentation_misc/bibtex/refs}

\end{document}